\documentclass[letterpaper,twocolumn,10pt]{article}

\usepackage[table]{xcolor}
\usepackage{float}
\usepackage{usenix2019_v3}
\usepackage{fancyhdr}
\usepackage{graphicx}
\usepackage{subfigure}
\usepackage{multirow}
\usepackage{amsmath}
\usepackage[ruled,vlined]{algorithm2e}
\usepackage{bm}
\usepackage{amsfonts}
\usepackage{url}
\usepackage{amsmath}
\usepackage{soul}
\usepackage{comment}
\usepackage{tabularx}
\usepackage{array}
\usepackage{rotating}
\usepackage{booktabs}
\usepackage{overpic}
\usepackage{float}
\usepackage{setspace}

	\newcommand{\ourtool}{BagAmmo}
	\newcommand{\ouralgorithm}{Apoem}

\begin{document}
\date{}

\title{\Large \bf Black-box Adversarial Example Attack towards {FCG} Based Android Malware \\ Detection under Incomplete Feature Information}

\author{Heng Li$^1$,
	 Zhang Cheng$^1,3$,
	 Bang Wu$^1$,
	 Liheng Yuan$^1$,
	 Cuiying Gao$^1$,
	 Wei Yuan$^1$,
	 Xiapu Luo$^2$ \\
	 $^1$ Huazhong University of Science and Technology\\
	 $^2$ The Hong Kong Polytechnic University\\
	 $^3$ NSFOCUS Technologies Group Co., Ltd\\
 }

\maketitle

\begin{abstract}
The {function call graph (FCG) } based Android malware detection methods have recently attracted increasing attention due to their promising performance. However, these methods are susceptible to adversarial examples (AEs). In this paper, we design a novel \textit{black-box} AE attack towards the {FCG} based malware detection system, called {\ourtool}. {To mislead its target system, {\ourtool} purposefully perturbs the FCG feature of malware through inserting "never-executed" function calls into malware code. The main challenges are two-fold. First, the malware functionality should not be changed by adversarial perturbation. Second, the information of the target system (e.g., the graph feature granularity and the output probabilities) is absent.}

{To preserve malware functionality, {\ourtool} employs the \textit{try-catch trap} to insert function calls to perturb the FCG of malware. Without the knowledge about feature granularity and output probabilities, {\ourtool} adopts the architecture of generative adversarial network (GAN), and leverages a multi-population co-evolution algorithm (i.e., {\ouralgorithm}) to generate the desired perturbation. Every population in {\ouralgorithm} represents a possible feature granularity, and the real feature granularity can be achieved when {\ouralgorithm} converges. }
	
{Through extensive experiments on over 44k Android apps and 32 target models, we evaluate the effectiveness, efficiency and resilience of {\ourtool}.  {\ourtool} achieves an average attack success rate of over 99.9\% on MaMaDroid, APIGraph and GCN, and still performs well in the scenario of concept drift and data imbalance.} {Moreover, {\ourtool} outperforms the state-of-the-art attack SRL in attack success rate.} 
	
	\end{abstract}
	
	\section{Introduction}

Occupying about $85\%$ of the global mobile operating system market, Android has become the main target of mobile malware in the world. A recent security report shows that on average, about 10000 new mobile malware samples were captured per day \cite{gdata}. The rapidly increasing of malware poses severe threats to Android users \cite{DBLP:conf/uss/SunSLM21,DBLP:conf/icssa/KimL18,DBLP:journals/tmc/LiuLZWZ20}, e.g., privacy leakage and economic losses. To tackle this problem, a variety of machine learning based Android malware detection methods have been designed to identify malware based on their features \cite{DBLP:conf/ccs/ZhangZZDCZZY20,DBLP:conf/ndss/MaricontiOACRS17,DBLP:conf/kbse/WuLZYZ019,DBLP:journals/tsmc/YuanJLC21,DBLP:conf/ndss/ArpSHGR14,DBLP:journals/tii/LiSYLSY18,DBLP:conf/kdd/HouYSA17,DBLP:journals/tifs/WangWFLHZ14,DBLP:conf/ccs/ZhangDYZ14,DBLP:conf/icccn/HuTMZZH14}. As a common feature for Android malware detection, Function Call Graph (FCG)  \cite{DBLP:conf/ccs/ZhangZZDCZZY20,DBLP:conf/ndss/MaricontiOACRS17,DBLP:conf/kbse/WuLZYZ019,DBLP:conf/kdd/HouYSA17,DBLP:journals/tifs/WangWFLHZ14,DBLP:conf/ccs/ZhangDYZ14,DBLP:conf/icccn/HuTMZZH14} (e.g., frequent subgraph \cite{DBLP:journals/tifs/Fan0LCTZL18} and E-FCG \cite{DBLP:journals/ijon/CaiJGLY21}) provides important clues for understanding how Android apps work. {In an FCG}, every node represents a function or an \textit{abstracted} function (e.g., class, package or family), and every edge denotes the calling relationship between caller and callee. 
\begin{figure}[bhtp]   
	\centering   
	\includegraphics[width=1\linewidth]{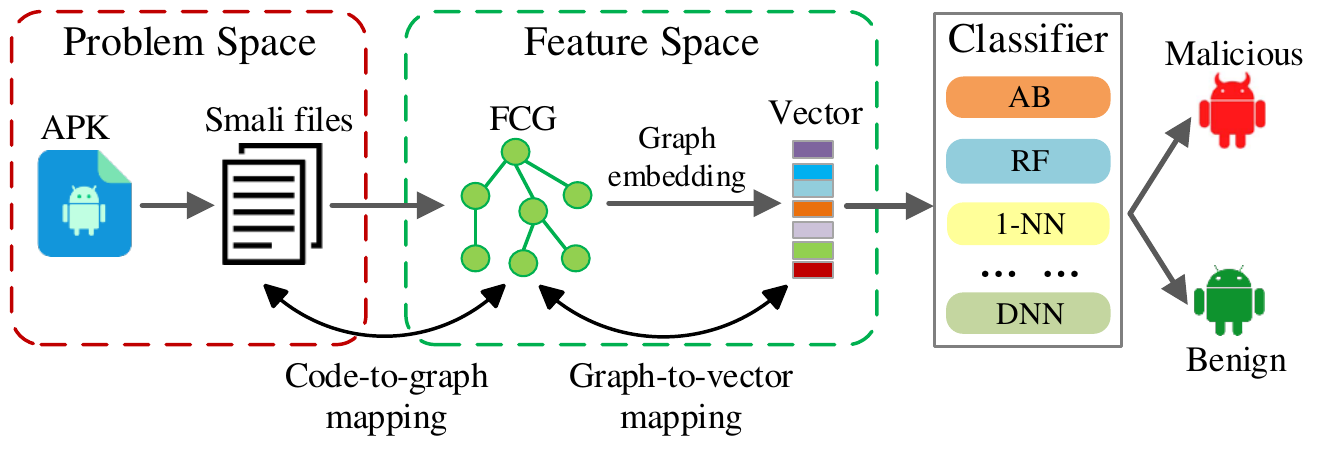}
	\caption{{FCG based Android malware detection framework.} }  
	\label{fig:motivation}   
\end{figure} 
As depicted in Fig. \ref{fig:motivation}, {the FCG} based Android malware detection usually consists of three steps. First, the {FCG} feature (e.g., frequent subgraph) is extracted from the Android Package (APK) file. Second, the {FCG} is transformed into a feature vector, i.e., graph embedding. Third, the feature vector is processed for malware prediction. Existing studies \cite{DBLP:conf/ccs/ZhangZZDCZZY20,DBLP:conf/ndss/MaricontiOACRS17,DBLP:conf/kbse/WuLZYZ019} demonstrate that the {FCG} based Android malware detection methods can achieve promising performance. 

Unfortunately, the {FCG} based malware detection is susceptible to {adversarial examples (AEs) \cite{DBLP:journals/corr/SzegedyZSBEGF13,DBLP:conf/uss/SuyaC0020,DBLP:conf/sp/Carlini017,DBLP:conf/ijcai/XuC0CWHL19,DBLP:conf/www/SunWTHH20,DBLP:conf/kdd/0001WDWT21}}, which are generated by imposing well-crafted adversarial perturbations on normal examples to induce misclassification. To evade detection, an adversary just needs to manipulate a malicious app  by elaborately modifying (e.g., inserting non-functional function calls) and repackaging its code. Although malware manipulation takes place in problem space (depicted by the {first} box in Fig. \ref{fig:motivation}), it changes the {FCG} (e.g., adding new edges) and perturbs the feature vector in feature space (described by the {second} box in Fig. \ref{fig:motivation}). Once the perturbation helps the feature vector stride over the target classifier's decision boundary, the repackaged malware will evade detection. {Up to now, a variety of AE attacks towards Android malware detection have been proposed to produce evasive Android malware}. Most of them \cite{huang2018adversarial,grosse2017adversarial,hu2017generating,2019lh,DBLP:journals/tifs/LiL20} direct at non-graph features (i.e., syntax features) based detection models that use binary feature vectors  for app classification. 
 {Recently, increasing attention has been paid to} the AE attacks towards graph feature  (i.e., semantic feature) based detection models \cite{DBLP:journals/corr/abs-2110-03301}\cite{chen2020android}. {For example,  Bostani \emph{et al.} \cite{DBLP:journals/corr/abs-2110-03301} leverage random search to find optimal perturbation for APK files in a black-box setting. Chen \emph{et al.} \cite{chen2020android} propose a method to exert optimal perturbations on Android APK files.}

{Up to now, how to produce Android malware to circumvent the FCG based  detection is still an open issue. This motivates us to investigate the generation of AEs to fight against the FCG based Android malware detection. In practice, building evasive malware needs to consider the following realistic problems that have not been well addressed.}

\noindent (1) \textit{Malware functionality preservation}. {The malware manipulation should be able to} mislead its target classifier {in the premise of malware functionality preservation}. 

\noindent (2) \textit{Problem-feature space gap}. 
{Since the feature vector in feature space cannot be directly perturbed, adversaries have to modify malware code in problem space} and expect their modification brings about the desired adversarial perturbation on feature vector. 

\noindent(3) \textit{Strict black-box setting}. For adversaries, the target classifier is a strict black box and its architecture, parameters and output probabilities are all unknown. 

\noindent(4) \textit{{Feature information absence}}. Adversaries cannot get the {feature used by their target classifier, i.e., the FCG and the feature vector obtained by graph embedding (denoted in the second box of Fig. \ref{fig:motivation}).} {Moreover, a detection system may use one of several possible feature granularities, e.g., class level, package level and family level (as discussed in Subsection \ref{subsec: features}). In practice, the feature granularity information is often unavailable to adversaries. }


{To overcome the above challenges}, we design a \textbf{b}lack-box \textbf{a}ttacks towards {FC\textbf{G}} based \textbf{A}ndroid \textbf{m}alware detection with \textbf{m}ulti-population co-ev\textbf{o}lution, termed {\ourtool}. {{\ourtool} works under the \textit{incomplete feature information} condition, which means adversaries do not know the granularity of the FCG feature used by their target system. }
Our main tasks include designing a malware manipulation technique used in problem space, and developing an algorithm to derive adversarial perturbation in feature space. {{\ourtool} constructs a dedicated Generative Adversarial Network (GAN) and employs its \textit{generator}} to generate candidate manipulations under the guidance of its \textit{discriminator}. The generator is implemented by our proposed \textbf{A}dversarial multi-\textbf{p}opulation c\textbf{o}-\textbf{e}volution algorith\textbf{m} ({\ouralgorithm}). {\ourtool} iteratively queries its target detection system with manipulated samples, and gradually learns the desired manipulation from a sequence of query-reply pairs. {{\ourtool} uses the following techniques to overcome the above challenges.}

\noindent(1) {\ourtool} leverages a novel malware manipulation method "\textit{try-catch trap}" to insert never-executed function calls into malware code {for functionality preservation}.

\noindent(2) {{\ourtool} maps the FCG into a feature vector, which transfers the impacts of malware manipulation into feature space and hence bridges the problem-feature space gap.}

\noindent(3) To overcome the challenge of strict black box, the discriminator substitutes the target {classifier} and guides the generator to figure out the desirable manipulation rapidly.

\noindent(4) {In {\ouralgorithm}, every population corresponds to a possible feature granularity.} Owning to the cooperative evolution among populations, {\ouralgorithm} converges to the real feature granularity under incomplete feature information.

Our main contributions are summarized as follows.

\noindent$\bullet$ We propose a novel black-box AE attack {\ourtool} towards the {FCG} based Android malware detection. {{\ourtool} does not require  complete information about feature space, and hence it is a broad-spectrum attack with strong generalizability. }

\noindent$\bullet$  We  theoretically analyze why {\ouralgorithm} can mitigate the prematurity problem {that often plagues the evolution algorithms}. 

\noindent$\bullet$ {We conduct extensive experiments on three state-of-the-art (SOTA) malware detection methods MaMaDroid \cite{DBLP:conf/ndss/MaricontiOACRS17},  APIGraph    \cite{DBLP:conf/ccs/ZhangZZDCZZY20} and GCN\cite{DBLP:journals/corr/abs-2009-05602} with five classifiers (e.g., RF and DNN) under three feature granularities. {\ourtool} surpasses 
{the SOTA attack (i.e., reinforcement learning based method SRL) in our experiments.}} It achieves an average attack success rate of over
\textbf{99.9\%} on all \textbf{32} target detection systems. Our experiments also confirm the {\ourtool}'s attack efficiency and resilience to concept drift and data imbalance.


{\textbf{Roadmap.} The remainder of the paper is organized as follows: §\ref{subsec: Preliminaries} introduces preliminaries; §\ref{subsec: Problem} presents the problem formulation; §\ref{sec:Adversarial manipulation design} discusses how to manipulate  the malware; §\ref{sec:method} describes the algorithm of perturbation generation; §\ref{SEC:EXP} gives the performance evaluation; §\ref{sec:Related} reviews relevant work; §\ref{sec:Limitations} provides {the limitations and discussion.}}

	\section{Preliminaries} \label{subsec: Preliminaries}

\subsection{Features for Android malware detection}\label{subsec: features}

 {In this subsection, we focus on the static features that are obtained prior to app execution and widely used in Android malware detection.} Earlier studies devote more attention to syntax features, e.g., requested permissions \cite{DBLP:journals/tii/LiSYLSY18,DBLP:conf/ndss/ZhouWZJ12,DBLP:conf/ccs/EnckOM09} , intent actions \cite{DBLP:journals/tsmc/YuanJLC21,DBLP:conf/uss/OcteauMJBBKT13,DBLP:conf/ntms/FereidooniCYS16}, Inter-Component 
Communications (ICCs) \cite{DBLP:conf/ndss/FengBMDA17,DBLP:conf/icse/BaiX00M20} and API calls \cite{DBLP:conf/ndss/ArpSHGR14,DBLP:journals/jnca/Seo0SBY14}. Recently, semantic features \cite{DBLP:conf/ccs/ZhangZZDCZZY20,DBLP:conf/ndss/MaricontiOACRS17,DBLP:conf/kbse/WuLZYZ019} (e.g., FCGs) have attracted increasing attention. They can characterize the behavior and functionality of apps, and hence achieve promising performance.

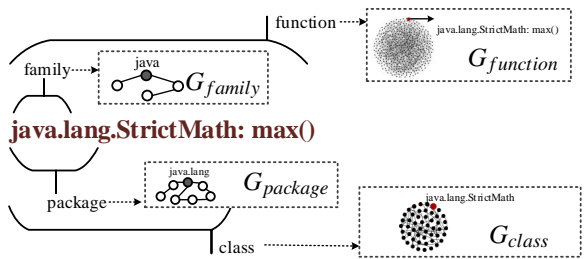
\begin{figure}[bhtp]   
	\centering   
	\includegraphics[width=0.9\linewidth]{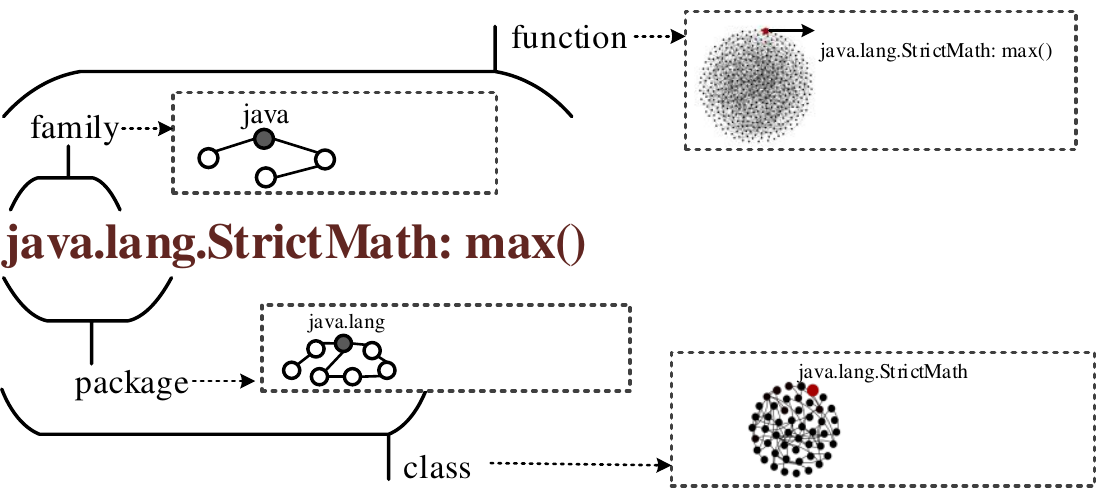}
	\put(-150,66){${G}_{family}$}
    \put(-128,28){${G}_{package}$}
	\put(-43,75){${G}_{function}$}
    \put(-35,8){${G}_{class}$}
	\caption{Different granularities of the FCG.}   
	\label{fig:granularities}   
\end{figure}

{ As the most common semantic feature, FCGs are often constructed based on smali files}. { A function or an \textit{abstracted} function denoted by its function name (e.g., java.lang.StrictMath: max()), class name (e.g., java.lang.StrictMath), package name (e.g., java.lang), or family name (e.g., java) can be used to represent a node in an FCG.}  Therefore, there exist four feature granularities in FCGs , i.e., function level, class level, package level and family level, as shown in Fig. \ref{fig:granularities}. 
{The features with finer granularities (e.g., class level)  usually have a more complex graph structure, causing heavier computational overhead and requiring dimensionality reduction \cite{DBLP:conf/ndss/MaricontiOACRS17}.}

 
{Clearly, the knowledge about the feature granularity of the target system is helpful for adversaries to generate AEs. However, this prior knowledge is hard to obtain in practice. Hence, we put forward the incomplete feature information assumption, assuming that adversaries do not know the feature granularity of the target system. }

\subsection{{FCG} based Android malware detection}



Here we introduce three state-of-the-art {FCG} based detection methods, which will act as the target detection systems in our experiments.  

\noindent \textbf{Mamadroid}. Mamadroid \cite{DBLP:conf/ndss/MaricontiOACRS17} considers the package-level or family-level FCGs as its features. More specifically, it adopts $340$ packages and $11$ families. To extract a feature vector from an FCG, Mamadroid constructs a Markov chain with the transition probabilities among packages or families. The extracted feature vectors are then used to train a classifier (e.g., KNN and SVM) for app classification. 


\noindent \textbf{APIGraph}. Different from Mamadroid, APIGraph \cite{DBLP:conf/ccs/ZhangZZDCZZY20} is a general framework {for} further enhancing the performance of the graph based Android malware detection methods. It employs a clustering algorithm (e.g., K-means) to aggregate the nodes (i.e., functions) of an FCG, based on the similarity among their semantics. It then uses a specific function to represent all functions in every cluster. Finally, APIGraph builds a new FCG with coarser granularity, in which every node denotes a cluster of functions and every edge indicates the call between two clusters. Experiments show that the new FCG can result in better classification performance. 


\noindent {
 \textbf{GCN}. 
Graph Convolutional Network (GCN) is a powerful graph embedding method, which can be utilized to detect malware. For instance, the GCN is used to convert the control flow graph into a feature vector for malware detection in \cite{DBLP:journals/corr/abs-2009-05602}\footnote{ {\cite{DBLP:journals/corr/abs-2009-05602} mainly studies how to attack malware detectors, although it proposes a GCN based malware detection method.} }. In Section \ref{SEC:EXP}, we will apply the GCN to the FCG based Android malware detection. } 

{ While these methods have achieved impressive results, they are susceptible to adversarial examples. The existence of adversarial examples is attributed to the problem that the decision boundaries of classification models are non-ideal \cite{DBLP:journals/corr/TanayG16,HU2022108824}. This problem becomes more serious in Android malware detection since the static analysis methods cannot precisely model the malware behavior. Therefore, the existing Android malware detection systems are not really secure\cite{277204}.}
	
\section{Problem formulation} \label{subsec: Problem}
 {Here we first introduce the system and threats considered in our work, and then propose an attack formulation to guide the design of black-box AE attacks.}
\subsection{{System \& Threat}}

\begin{figure}[htbp]
	\centering
	\includegraphics[scale=0.8]{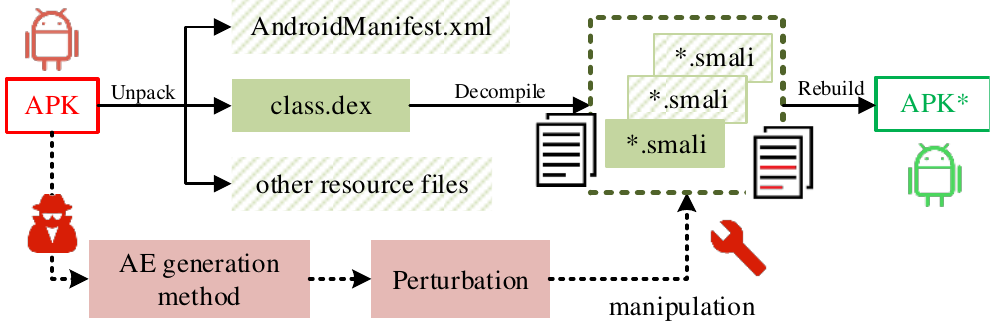}
	\caption{Overview of the AEs generation.}
	\label{fig:manipulation overview}
\end{figure}

{Fig. \ref{fig:motivation} depicts the {FCG} based Android malware detection system considered in this work. Suppose an adversary launches a black-box AE attack towards this system to produce real evasive malware. To this end, the adversary first gets the classes.dex file from an APK file, and further decompiles it into a series of smali files, as shown in Fig. \ref{fig:manipulation overview}. The adversary manipulates the smali code according to its perturbation, and rebuilds the code to obtain a new APK file. The adversary then queries the detection system with the generated malware sample, utilizes the received binary decision (i.e., benign or malicious) to update its perturbation, and then rebuilds a new malware sample. The above procedure is repeated until a real evasive malware is obtained. }

The adversary only knows that the target system uses FCG feature for malware detection. However, the adversary does not know the feature granularity and the graph embedding method used by the target system. Moreover, the adversary has no information about the architecture, the parameters and the output probabilities of the target classifier. As for the defender, it can use static analysis and white list based defenses to resist evasive malware. In addition, the defender may raise alarms once the number of queries from a user is unusually large\footnote{ {
Our experiments indicate that our method only needs dozens of queries to generate the perturbations that can successfully attack the target model. Moreover, our method can further reduce the number of perturbations by conducting more queries (e.g., several hundreds of queries). To accelerate the attack process, we provide a substitute network to fit the target model. The related experiments can be found in  Section \ref{sec:RQ2}}}.



\subsection{Attack formulation}
For convenience, {we first use $s$ and $m$ to refer to the malware sample and the manipulation}, respectively. We then use two functions $\mathcal M_G(\cdot)$ and $\mathcal M_V(\cdot)$ to denote the code-to-graph mapping and the graph-to-vector mapping shown in Fig. \ref{fig:motivation}, respectively. Through manipulating the malware sample $s$ with $m$, the adversary changes the input graph from $G=\mathcal M_G(s)$ to $\widetilde{G}=\mathcal M_G(s+m)$, where $G$ and $\widetilde{G}$ represent the original input {FCG} and the perturbed input {FCG}, respectively. Suppose $L(\cdot)$ denotes the label (i.e., benign or malicious) predicted by the target classifier. Then, the desired adversarial manipulation $m^*$ can be derived by solving the following problem:
\begin{equation}
L(\mathcal M_V(\mathcal M_G(s))) \neq L(\mathcal M_V(\mathcal M_G(s+m^*))) \label{objective: misleading}
\end{equation}
 {under the constraint of malware functionality preservation. }

{The above formulation points out two tasks for us: 1) designing a manipulation technique to modify malware code while preserving malware functionality, and 2) developing an adversarial perturbation generation algorithm to realize $m^*$. Due to the challenges of problem-feature space gap and strict black-box setting, $\mathcal M_G(\cdot)$ and $\mathcal M_V(\cdot)$ are actually unknown to the adversary. Hence it is extremely hard to derive the desired adversarial perturbation in one shot.} This motivates us to develop an evolutionary algorithm (i.e., {\ouralgorithm}) to gradually find the desired perturbation. {We will discuss how to fulfill  the above two tasks in Sections \ref{sec:Adversarial manipulation design} and \ref{sec:method}, respectively. } 

{Furthermore, it is noted that a variety of graph adversarial attack models \cite{DBLP:conf/ijcai/XuC0CWHL19,DBLP:conf/icml/BojchevskiG19,DBLP:conf/ijcai/Wu0TDLZ19,DBLP:conf/iclr/ZugnerG19,DBLP:conf/icml/DaiLTHWZS18,DBLP:conf/kdd/0001WDWT21,DBLP:conf/www/SunWTHH20} have been proposed in the community of machine learning. 
Although these methods offer inspirations to us, they cannot be directly applied to our attack for two reasons. First, graph adversarial attack models launch attacks from feature space. However, the attack against Android malware detection cannot directly access feature space, and has to indirectly affect feature space through manipulating malware code in problem space. Second, our attack needs to meet practical requirements (i.e., \textbf{R1}-\textbf{R4} discussed in Subsection \ref{subsec: four R}), which are absent in existing graph adversarial attacks. Therefore, specialized study is needed for malware adversarial attack design. }

\section{ Malware manipulation} \label{sec:Adversarial manipulation design}
{
In this section, we first introduce the common requirements and the existing techniques \cite{DBLP:conf/sp/PierazziPCC20,DBLP:journals/corr/abs-2009-05602,chen2020android} of malware manipulation, and then propose a new malware manipulation technique.}

\subsection{Background of malware manipulation} \label{subsec: four R}
{Although the manipulation on malware is intuitively simple, the challenges come from the following requirements.} \\
\textbf{R1: Functional Consistency.} The malware functionality should keep consistently before and after manipulation.\\
\textbf{R2: All-granularities influence.} Since the feature granularity (e.g., family level and package level) of malware detection is unknown, malware manipulation should be able to affect the features of all granularities \cite{DBLP:conf/ndss/MaricontiOACRS17}. \\
\textbf{R3: Resilience to static analysis.}  Malware manipulation 
 {should not be hindered by static analysis inspection}
\footnote{{ In this work, the static analysis mainly refers to the program analysis techniques that only examine the source code but do not execute the program.}} \cite{DBLP:journals/tdsc/DemontisMBMARCG19,DBLP:conf/acsac/MoserKK07}, and cannot completely rely on dead codes  {(i.e., unreachable instruction blocks)}.\\
\textbf{R4: Non-stationary perturbation.} Manipulation  should be non-stationary and cannot be restricted to a fixed set of operations (e.g., a pre-determined white list  \cite{DBLP:journals/corr/abs-2009-05602}), {to reduce the risk of being identified. } \\

{Existing manipulation methods are summarized below.}\\
\textbf{Inserting dead codes}: To maintain functional consistency, \cite{chen2020android} chooses to insert dead codes (e.g., no-op calls) into smali files. {Unfortunately, these codes can be easily detected and filtered, violating the requirement \textbf{R3}. For example, \cite{DBLP:journals/tifs/Fan0LCTZL18} proposes a weighted sensitive-API-call-based Android malware family classification method, which can resist the impact of no-op calls. } \\
\textbf{Adding valueless calls}: \cite{chen2020android} creates user-defined classes and adds valueless calls (i.e., invoking empty functions) into them. {However, these calls may be susceptible to static analysis and cannot attack the class-granularity FCGs, violating the requirements \textbf{R2} and \textbf{R3}.  For example, the Android malware detection method proposed in \cite{DBLP:journals/tsmc/YuanJLC21} does not use self-defined functions as feature. Hence, this method is not influenced by the valueless calls inserted by adversaries. }\\
\textbf{Adding functions from a white list}: To change FCGs, the authors of \cite{DBLP:journals/corr/abs-2009-05602} add a function coming from a predetermined white list. However, once adversarial examples are captured, the white list will be revealed and adversarial attacks may fail. Please refer to  requirement \textbf{R4}.\\
{\textbf{Opaque predicates}: \cite{DBLP:conf/sp/PierazziPCC20} leverages opaque predicates to insert new APIs for malware detection evasion. Specifically, this method constructs obfuscated conditions where the outcome is always known in design phase but the truth value is difficult or impossible to determine by static analysis. Hence this method can effectively resist static analysis. However, it may introduce some undesired functions (e.g., the \textit{random} function), which impose unexpected impacts on FCGs. }


\subsection{The proposed manipulation method} \label{sec:adversarial manipulation method}
Here we design a new malware manipulation method to modify smali code. Clearly, we cannot remove nodes or edges from FCGs, according to the requirement \textbf{R1}. Hence, we only consider adding (or inserting) nodes or edges. However, adding isolated nodes (i.e., the functions that are not invoked or do not invoke others) is not recommended for two reasons. First, the isolated nodes are easily detected by static analysis {(e.g., some program analysis techniques that perform redundant code elimination would remove unreachable code \cite{guardsquare})}. Second, adding nodes usually cannot impact feature space, since lots of malware detectors utilize edges (instead of nodes) for classification. As a result, we select adding edges (i.e., calls) in our manipulation method. Then the rest of the problem includes:  how to create \textit{candidate} edges,  how to select desirable edges from the candidate edges, and  how to insert the selected edges.  {
In this section, we only consider the first and the third problems. The second problem will be solved in Section \ref{sec:method}.}

\textbf{(1) How to create candidate edges?}
\begin{figure}
	\centering
	\includegraphics[scale=0.85]{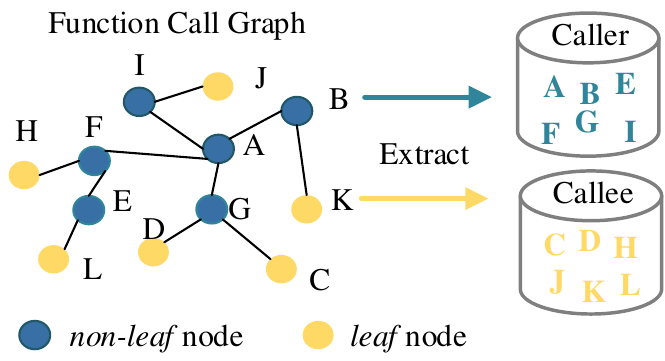}
	\caption{Selecting callers and callees from an FCG.}
	\label{fig:example}
\end{figure}

\begin{figure}
	\centering
	\includegraphics[scale=0.55]{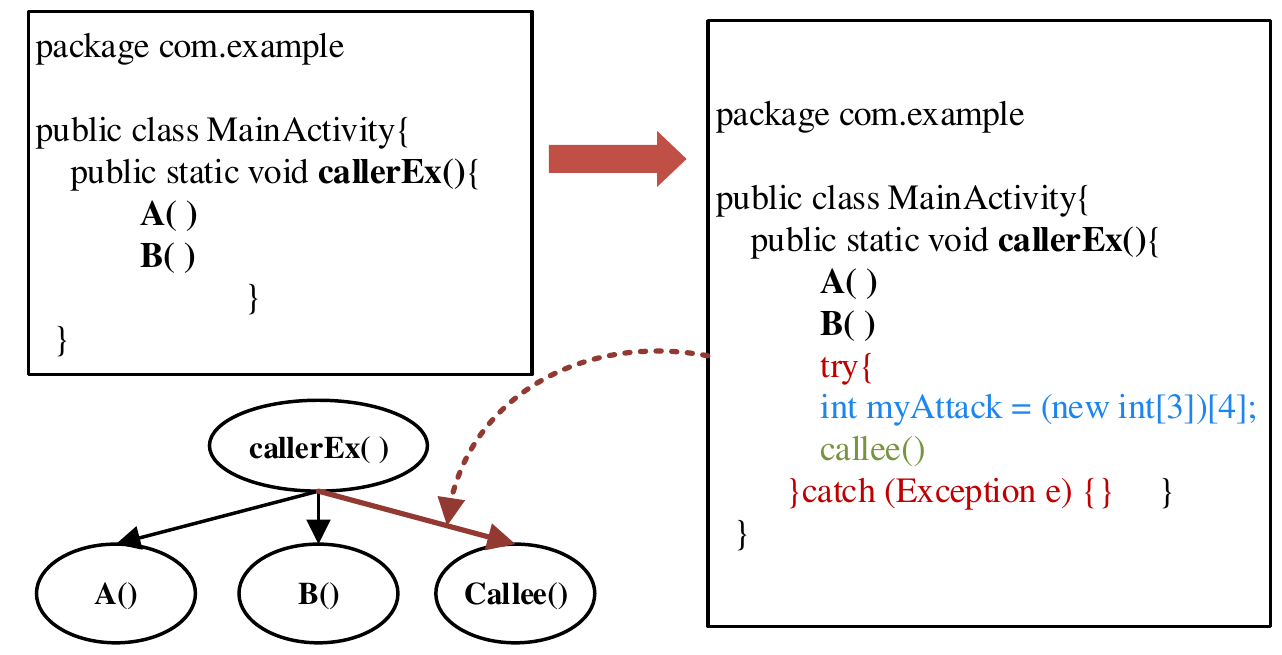}
	\caption{An example of try-catch trap.}
	\label{fig:manipulation}
\end{figure}

 {Up to now, how to impose all-granularities influence (required by \textbf{R2}) on FCG with incomplete feature information (i.e., the feature granularity is unknown) has not been thoroughly studied. To tackle this problem, we propose to create an edge between two nodes of any type by adding a function call between a caller and a callee. This method changes the FCG no matter what kind of feature granularity is used. } Then the problem becomes how to determine the caller and the callee for every candidate edge. Due to the requirement \textbf{R4}, we cannot utilize a white list to generate callers and callees. Instead, we propose to generate them from the functions used by malware itself. In this way, we can ensure that the candidate edges created for different malware are diverse, hence satisfying the requirement \textbf{R4}.  

{Now we study where to place the added edges.} An FCG consists of \textit{non-leaf} nodes and \textit{leaf} nodes, as depicted in Fig. \ref{fig:example}. The non-leaf nodes are user-defined functions, and the leaf nodes correspond to Android standard functions (e.g., $java/io/File;-\textgreater exists()$) or the user-defined functions that do not invoke others. In our method, non-leaf nodes (i.e., user-defined functions) are selected as callers, since they are easily inserted with new function calls. Leaf nodes are chosen as callees,   since invoking a function that does not invoke others will not trigger unintended calls. Here we avoid generating unintended calls because they may further impose perturbations on the FCG, which is beyond our expectation. Furthermore, we supply more discussion on callee selection in Appendix \ref{sec:callees' limitation}. Now we can use the above method to create candidate edges. In Section \ref{sec:method}, we will propose an algorithm to select the most desirable edges for manipulation.

\textbf{(2) How to insert selected edges?}

 We assume that the desirable edges have been selected, and study how to insert the corresponding function calls into smali files under the requirements of \textbf{R1} and \textbf{R3}. Our proposed method is called \textbf{try-catch trap}. It first inserts a try-catch block into the caller, and places the statement of invoking callee in its try block. It then adds several statements in front of this function call statement. These statements are used to trigger a pre-selected exception (e.g., an arithmetic exception). Now we analyze why this method works. First, it inserts a function call statement in smali files, hence changing the FCG by adding a new edge. Second, the function call statement is never executed,  {hence preserving malware functionality}. For illustration, Fig. \ref{fig:manipulation} gives an example of try-catch trap. Suppose the codes in the left box come from a malware sample. The function \textit{callerEX()} is selected as our caller. We place a try-catch block in this function, and invoke the function callee() after the blue statement is executed. In this way, we can add a new edge into the FCG, as shown in Fig. \ref{fig:manipulation}. When the try-catch block is executed, an exception of \textit{IndexOutOfBoundsException} will be thrown, and the statement of function call will be skipped over. 
{In summary, our method can be considered as a variant of opaque predicates. It carefully constructs obfuscated conditions that are difficult to determine during static analysis, hence possessing the ability to resist static analysis.}

 {The main steps of inserting function calls are briefly described in Appendix \ref{appendix:smali}.}

\newcommand{\tabincell}[2]{\begin{tabular}{@{}#1@{}}#2\end{tabular}}


\section{Adversarial perturbation generation} \label{sec:method}
{ In Subsection \ref{sec:adversarial manipulation method}, we propose the question of how to
 select desirable edges from the candidate edges. To answer this question, we develop a novel GAN model and the algorithm {\ouralgorithm} to find the desired adversarial perturbation. } 

\subsection{Challenges \& Solutions}

{We first introduce the main procedure of {\ourtool} below. }

 (1) Given a pre-selected malware sample, {\ourtool} finds some callers and callees from the smali codes, and uses them to create a set of candidate edges, as discussed in Section \ref{sec:Adversarial manipulation design}. With candidate edges, {\ourtool} generates a variety of samples through manipulating malware, and sends them (i.e., queries) to its target system for malware detection.
 
 (2) The target model sends back a reply for a query. In our strict black-box setting \cite{DBLP:conf/iclr/ZhaoDS18}, every reply contains only the binary classification outcome (i.e., malicious or benign). 
 
 (3) Through learning from the query-reply pairs, {\ourtool} gradually recognizes the most desirable edges that can successfully induce misclassification.

 The main challenges in designing {\ourtool} include: 1) {the feature granularity} of the target model is unknown, 2) a large number of queries are usually required in the strict black-box attack scenario \footnote{{In this scenario, both $\mathcal M_G(\cdot)$ and $\mathcal M_V(\cdot)$ mentioned in Subsection 3.2 are unknown. Moreover, the reply of the   black-box model contains only the
binary classification outcome (e.g., many malware detection websites \cite{VirusTotal} only sends back a binary decision instead of class probabilities). }} \cite{DBLP:conf/icdm/LiYZ18,DBLP:conf/eccv/AndriushchenkoC20}.
Our countermeasures are briefly explained below. 

\textbf{Surmising feature granularity}. {Our adversarial multi-population co-evolution algorithm, i.e., {\ouralgorithm}, uses one population to represent a possible feature granularity}. {The multiple populations, corresponding to multiple possible feature granularities}, cooperatively evolve until the population corresponding to the real feature granularity keeps alive but the others fade away. In this way, {\ourtool} can accurately identify the feature granularity used by its target model, {as will be shown in Subsection \ref{subsection:evolution}}. 

\textbf{Reducing the number of queries}. {\ourtool} constructs a novel \textit{substitute} model to simulate its target model. The substitute model is trained with the samples generated by {\ouralgorithm} and labeled by the target model. {As will be shown in Subsection \ref{subsection: substitute}}, once the substitute model is well trained, {\ourtool} only needs to attack it instead of the target model, {hence greatly reducing} the number of queries. 

\subsection{The overview of {\ourtool}}
{Following the architecture of GANs, {\ourtool} adopts a generator and a discriminator that are cooperatively trained.} 

\textbf{Generator}: The generator is responsible for generating perturbations, i.e., the new edges added into the FCG. It is implemented with an adversarial multi-population co-evolution algorithm (i.e., Apoem).

\textbf{Discriminator}: {The discriminator is introduced to stimulate the generator to improve its perturbations. It is implemented with a GCN, acting as a substitute network \cite{chen2020android} to simulate the target model.}


\textbf{Training}: {In each round of model training}, the generator modifies the malware's code and sends the rebuilt malware to the target model or the substitute model for malware detection. {\ourtool} makes a choice between the target model and the substitute model with a variable probability $p$. After receiving the queries, the target model sends back its replies, i.e., the binary decisions. With the query-reply pairs, {\ourtool} trains the substitute model and guides its generator to {improve its generated perturbations}. {The probability $p$ keeps growing as the number of rounds increases, to decrease the number of queries sent to the target model}.

\begin{figure*}[htbp]
	\centering
	\includegraphics[scale=0.5]{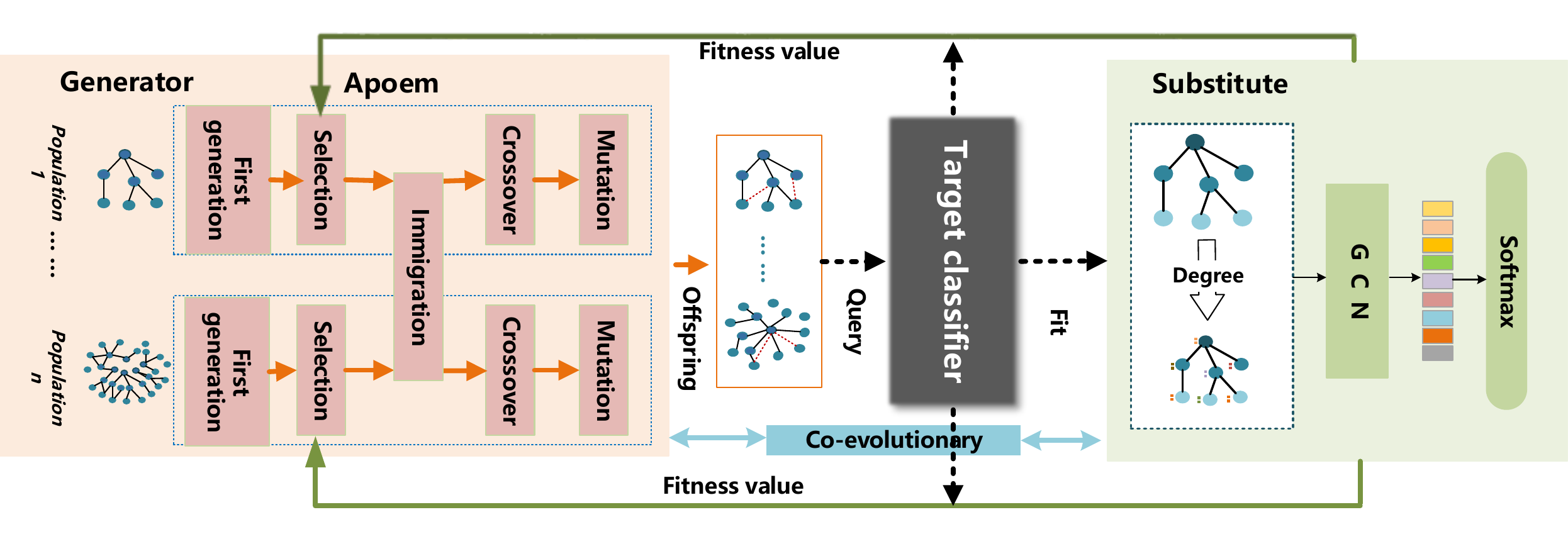}
	\caption{The model architecture of {\ourtool}.}
	\label{fig:bg1}
\end{figure*}

\subsection{Adversarial Multi-population co-evolution}
\label{subsection:evolution}
\begin{figure}[htbp]
	\centering
	\includegraphics[scale=0.6]{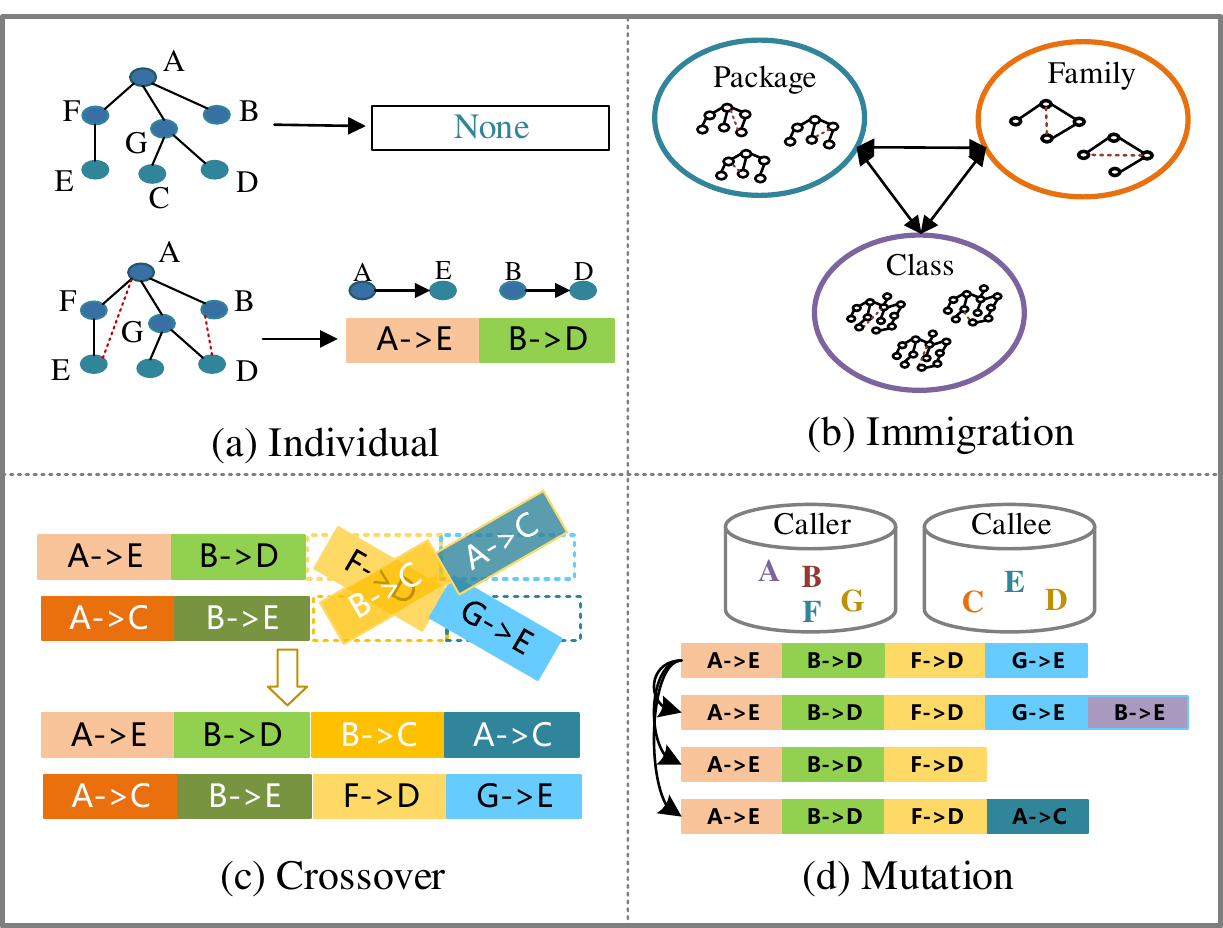}
	\caption{{How multiple populations cooperatively evolve?}}
	\label{fig:GA}
\end{figure}
    { The main challenge faced by the generator is 
  {that the real feature granularity is unknown.} To facilitate the understanding, we consider the case where the target system uses the family-level feature but we perturb the class-level feature. In this case, we will fall into a huge search space, hence prolonging model training time and requiring more queries. To alleviate this problem, {\ourtool} uses the {\ouralgorithm} algorithm to surmise the real feature granularity.} {\ouralgorithm} follows the general framework of evolutionary algorithms, but it introduces  cooperation among multiple populations to speed up convergence. Along with the evolution, the population corresponding to the real feature granularity gradually stands out from the crowd. In the following, we first describe the main components of {\ouralgorithm} depicted in the red block of Fig. \ref{fig:bg1}, and then discuss how to use these components {to generate the desired perturbation}. 
 
\textbf{(1) Population \& Individual}. 
{A population represents  a collection of generated AEs under a certain feature granularity. For example, the family-level population consists of the AEs generated under the assumption that the target classifier uses a family-level FCG as its input. {\ouralgorithm} adopts multiple populations, each of which corresponds to one possible feature granularity (i.e., family, package and class)}.
Each individual in a population gives a perturbation that can be imposed on the original FCG\footnote{Strictly speaking, an individual refers to an adversarial example in a population. However, the difference between adversarial example and malicious example is perturbation. Hence we use the perturbation to represent an individual.}, i.e., the set of edges added into the FCG. As shown in Fig. \ref{fig:GA} (a), the above graph denotes the original FCG, and the below graph represents an adversarial example. Accordingly, the perturbation, i.e., the edge set $(A\rightarrow E, B \rightarrow D)$, is considered as an individual. {We use $x_r^{(i,j)}$ to refer to the $j$-th individual of the $i$-th population in the $r$-th generation of {\ouralgorithm}. We have $x_r^{(i,j)} = \{ e_1^{(i,j)},e_2^{(i,j)},...,e_{n}^{(i,j)}\}$, where $e_{k}^{(i,j)}$ ($1\leq k\leq n$) is the added edge.}
	In the initial phase, we need to collect sufficient individuals to build the populations. Therefore, we randomly perturb the original FCG, and get a set of individuals for each population.
	
 \textbf{(2) Fitness \& Selection}.
	{{\ouralgorithm} employs the metric fitness to select superior individuals and eliminate inferior individuals.}
	{This metric reflects the aggressivity and the invisibility of an AE. Its calculation takes into account two factors:} threat degree $T$ and perturbation amount $L$. The threat degree is measured according to the output of the target model $F(\cdot)$ or the substitute model $S(\cdot)$\footnote{{For the target or substitute model, the input is detected to be malicious when its output $F(x)$ or $S(x)$ equals to or approaches 1.  }}. For an individual $x$, the threat degree is defined as:
		{
	\begin{equation}
	 T =\left\{\begin{array}{l}
1-F(x) \quad\text{if target model is used} \\
1-S(x) \quad\text{if substitute model is used}
\end{array}\right. 	    
	\end{equation}}
The perturbation amount is calculated as the number of added edges.
	{Furthermore, {\ouralgorithm} introduces the \textit{elitist} selection strategy \cite{2012Genetic}} to pass on the good genes of individuals to the next generation, { through retaining the fittest individuals and eliminating the others.} 

\textbf{ (3) Immigration}. {In general, the individuals with high fitness have a greater chance of producing better offsprings. To produce more high-quality individuals, {\ouralgorithm} leverages the immigration operation to transfer individuals with high fitness within one population into other populations. Accordingly, the superior individuals immigrate to different populations, making all populations cooperatively evolve to generate better AEs.}
{There exist two kinds of immigration in {\ouralgorithm}: fine-to-coarse (e.g., from class level to family level) and coarse-to-fine (e.g., from family level to class level), as shown in Fig. \ref{fig:GA} (b). We first consider the fine-to-coarse case where one individual in the class-level population is immigrated into the package-level population. In this case, the name of the packages related to the perturbation (e.g., java.lang.StrictMath->java.lang) is retained and the individual containing only package names is then put into the package-level population. Now we consider the coarse-to-fine case where the individual from the package-level population is injected into the class-level population. Since a package may contain multiple classes, we randomly select one class used by malware code to replace the package and then put the individual containing class names into the class-level population.}

\textbf{(4) Crossover}. 
{{\ouralgorithm} leverages crossover to randomly swap genes from two parents to produce offsprings. More specifically, $K$ pairs of individuals are randomly chosen from a population as parents, and half of the perturbation in every pair is exchanged to produce two offsprings, as shown in Fig. \ref{fig:GA} (c).} { Suppose the parents are $	 x_r^{(i,j_1)} =\{e_1^{(i,j_1)},e_2^{(i,j_1)},e_3^{(i,j_1)},e_4^{(i,j_1)}\}$ and $ x_r^{(i,j_2)} = \{e_1^{(i,j_2)},e_2^{(i,j_2)},e_3^{(i,j_2)},e_4^{(i,j_2)}\}$, where  $e_{k}^{(i,j)}$ is an added edge (e.g., A->E)  in Fig. \ref{fig:GA} (c) . The offsprings derived by crossover are $x_{r+1}^{(i,j_1)} = \{e_1^{(i,j_1)},e_2^{(i,j_1)},e_3^{(i,j_2)},e_4^{(i,j_2)}\}$ and $x_{r+1}^{(i,j_2)} = \{e_1^{(i,j_2)},e_2^{(i,j_2)},,e_3^{(i,j_1)},e_4^{(i,j_1)}\}$, respectively. }

 \textbf{(5) Mutation}. 
{{\ouralgorithm} employs mutation to bring new changes to a population.} As depicted in Fig.  \ref{fig:GA} (d), there are three possible mutation modes: 1) randomly adding function calls on the existing perturbation, 2) randomly reducing existing perturbation, and 3) randomly exchanging existing  perturbations. {They can be mathematically expressed as $x_{r+1}^{(i,j)}=\{e_1^{(i,j)},  ... ,e_{n}^{(i,j)},e_{n+1}^{(i,j)}\}$, $	x_{r+1}^{(i,j)}=\{e_1^{(i,j)},  ... ,e_{n-1}^{(i,j)}\}$, and $x_{r+1}^{(i,j)}=\{e_1^{(i,j)},  ... ,e_{n-1}^{(i,j)},e_{n+1}^{(i,j)}\}$, respectively.}

	 

\subsection{Substitute model}
\label{subsection: substitute}
{{\ouralgorithm} only knows the binary decision of its target model, making it hard to accurately evaluate individuals. To overcome this challenge, we design a novel substitute model to simulate the target model, and provide {\ouralgorithm} with approximate class probabilities.}

The inputs of our substitute model are \textit{function-level} FCGs generated according to the perturbation produced by the generator. 
{We use a GCN  {(i.e., Graph Convolutional Network)} to extract features from the substitute model}, as shown in the green block in  Fig. \ref{fig:bg1}. GCNs extend convolution to graph data, and they are good at utilizing structural information and node information to fulfill graph-related machine learning tasks. However, the main obstacle of applying GCNs to our task is the absence of node property. That is, FCGs do not provide property information for their nodes. To alleviate this problem, we propose to use \textbf{out degree} and \textbf{in degree} of a node as its features.

Now we briefly explain {how to use a GCN to extract} features from the inputs. The GCN has multiple convolutional layers. Each layer aggregates node properties using a propagation rule, and the aggregated features are then processed by the next layer. {Accordingly, we can obtain a feature vector to represent the FCG using iterative computation. 





\subsection{Algorithm design} 
{{\ouralgorithm} aims to conglutinate multi-population co-evolution mechanism and substitute model to cooperatively generate adversarial perturbations. Its main procedure is given in Algorithm 1.} In this algorithm, $F$ is the target classifier, $N$ is the maximum number of individuals in a population, and $r_ {max}$ denotes the maximum number of generations. {In every iteration, {\ouralgorithm} first randomly selects the target or the substitute model (lines 3-6),  calculates fitness for every individual (lines 8-12), then retains high-rate individuals based on fitness (line 13), and finally conducts immigration, crossover and mutation (line 14). In addition, the substitute model should be trained when it is selected, as denoted by lines 15-17. The individual with the highest fitness is outputted when {\ouralgorithm} terminates. Below we summarize the important considerations for {\ouralgorithm}}.
\begin{algorithm}[h]
	    \LinesNumbered
	\caption{The {\ouralgorithm} Algorithm}
	\label{algorithm ga}
	 \KwIn{The FCG $G$ of a given APP}



		 
		 
    Population initialization;

	
	\For{$r$ in $r_ {max}$ }{
		\uIf {$ (r-3)/r_ {max} > random(0, 1)$ }{
			Is\_substitute = 1;	}  
		\Else{Is\_substitute = 0;}

		\For{each $P^i$ }{
			\uIf {Is\_substitute = 1 }{Get fitness $T(x_r^{(i,j)})$ from substitute model;	}
			\Else{Get fitness $T(x_r^{(i,j)})$ from target model;}	
							
		 	Get  $L(x_r^{(i,j)})$ for every individuals;
		 
	     	Select top $N$ individuals according to  $T(x_r^{(i,j)})$ and $L(x_r^{(i,j)})$ in turn ; 
	     	
	     	Immigration(); Crossover();	Mutation();
	}
	
	\If {Is\_substitute = 1 }{
		Get the result from the target model $F(x_r^{(i,j)})$;
		
		Train substitute model with $F(x_r^{(i,j)})$ and $x_r^{(i,j)}$;}	
	
        Determine whether algorithm should terminates;
		
}


\end{algorithm}


\noindent (1) \textit{{How to implement co-evolution in {\ouralgorithm}?}}\\
{ The co-evolution in {\ouralgorithm} is two-fold. On one hand, the generator and the discriminator cooperate with each other to improve the generated perturbation. On the other hand, multiple populations cooperatively evolve through immigration.}



\noindent (2) \textit{How to avoid premature convergence?}\\
{ When the genes of some high-rate individuals quickly dominate the population \cite{1997Degree}, premature convergence occurs and evolutionary algorithms converge to a local optimum. {\ouralgorithm} can mitigate premature convergence owing to the cooperation among populations.} Through immigration, different populations share their good genes and further promote their evolution. Meanwhile, immigration also helps the populations jump out from local optimum traps. {Our  theoretical analyses are given in Appendix \ref{appendix: theoretical analyses}}.

\noindent (3) \textit{When to terminate our algorithm?}\\
There are three stopping criteria for {\ouralgorithm}. First, all the offsprings cannot induce misclassification on the target model anymore. Second, the perturbation amount does not decrease within several continuous rounds. Third, the maximum number of rounds is reached. 

\noindent (4)  {\textit{How to modify the APK according to the output?}\\
The output of our algorithm is the caller-callee function pairs. According to the output, we use the try-catch trap mentioned in Section \ref{sec:Adversarial manipulation design} to insert the callee function into the caller function, in order to implement adversarial perturbation. The  implementation details can be found in Appendix \ref{appendix:smali}.}

	
\section{Experiments}\label{SEC:EXP}

In this section, we conduct extensive experiments to evaluate {\ourtool} by answering the following research questions:\\
\textbf{RQ1: Effectiveness.} { Does {\ourtool} successfully attack  the SOTA Android malware detection methods? }\\
\textbf{RQ2: Evolution.} How do multiple populations in {\ouralgorithm} evolve?\\
\textbf{RQ3: Efficiency.} { Does the substitute model help to decrease
queries and improve attack efficiency? }\\
\textbf{RQ4: Overhead.} Is there a trade-off between manipulation overhead and attack success rate?\\
\textbf{RQ5: Resilience.} {Is {\ourtool} still effective when there exists concept drift or data imbalance?}\\
\textbf{RQ6: Functionality.} {Does our adversarial perturbation change the functionality of malware?}

\textbf{Datasets}.
{ Our dataset contains 21399 benign samples and 22975 malicious samples, which come from Androzoo\footnote{https://androzoo.uni.lu/}, Faldroid dataset \cite{DBLP:journals/tifs/Fan0LCTZL18} and Drebin dataset \cite{DBLP:conf/ndss/ArpSHGR14}. Every sample collected from  Androzoo is detected by VirusTotal \cite{VirusTotal}. Only when a sample is detected to be malicious by more than four antivirus systems, we label it as malware. The details of our dataset  are provided in Appendix \ref{appendix: dataset}.} 

Furthermore, our experiments adopt two configurations to evaluate {\ourtool}. According to the first configuration,  we use 10-fold cross-validation to train the target models. To evaluate the attack methods, we randomly choose $100$ malicious examples (not included in the training data of target model) that can be correctly classified by target models for evasive malware generation. For the second, we divide the dataset according to the years that Android apps emerge as discussed in Section \ref{subsection: resilience}. The newly-emerged malware samples are used for test, while the old data are used in training.




 {Finally, we also consider the scenarios of concept drift and data imbalance in Subsection \ref{subsection: resilience}. In the scenario of concept drift, 17685 samples (8,017 benign examples and 9,668 malicious examples) from Androzoo are grouped by production year (from 2016 to 2020) and used to train target models. In the scenario of data imbalance, we randomly disarrange samples and set the benign-malicious ratio to 10:1, following the experimental setting in \cite{277204}.     }


\begin{table*}[!ht]
	\caption{ {Effectiveness of {\ourtool} towards MaMaDroid, APIGraph and GCN.}}
	\centering
	\scalebox{1}{
		\begin{tabular}{c|c|ccc|ccc|ccc}
			\hline
			\multicolumn{2}{c|}{\multirow{2}{*}{Classifier\textbackslash{}Level}} & \multicolumn{3}{c|}{Family} & \multicolumn{3}{c|}{Package} & \multicolumn{3}{c}{Class} \\ \cline{3-11} 
			\multicolumn{2}{c|}{}                                                       & ASR     & APR     & IR      & ASR     & APR     & IR       & ASR    & APR    & IR      \\ \hline
			\multirow{5}{*}{MaMaDroid}                      
			& \cellcolor{cyan!60!gray!10}RF                        & \cellcolor{cyan!60!gray!10}1.000    & \cellcolor{cyan!60!gray!10}0.021    & \cellcolor{cyan!60!gray!10}8.670    & \cellcolor{cyan!60!gray!10}1.000    & \cellcolor{cyan!60!gray!10}0.049    & \cellcolor{cyan!60!gray!10}13.640    & \cellcolor{cyan!60!gray!10} 1.000  & \cellcolor{cyan!60!gray!10}	0.083  & \cellcolor{cyan!60!gray!10}	12.490    \\   

			& DNN                       & 0.990    & 0.149    & 11.130    & 1.000    & 0.134    & 16.730    & 1.000   & 	0.153  & 15.907    \\  	

			& \cellcolor{cyan!60!gray!10}AB                        & \cellcolor{cyan!60!gray!10}1.000   & \cellcolor{cyan!60!gray!10}0.066    & \cellcolor{cyan!60!gray!10}10.270    & \cellcolor{cyan!60!gray!10}1.000    & \cellcolor{cyan!60!gray!10}0.072    & \cellcolor{cyan!60!gray!10}14.300    & \cellcolor{cyan!60!gray!10}1.000  & \cellcolor{cyan!60!gray!10}0.118  & \cellcolor{cyan!60!gray!10}15.460   \\     		

			& 1-NN                      & 1.000    & 0.031    & 7.000    & 1.000    & 0.109    & 11.630     &1.000   & 0.060  & 10.960    \\  	 	

			& \cellcolor{cyan!60!gray!10}3-NN                      & \cellcolor{cyan!60!gray!10}1.000   & \cellcolor{cyan!60!gray!10}0.037    & \cellcolor{cyan!60!gray!10}9.390    & \cellcolor{cyan!60!gray!10}1.000    & \cellcolor{cyan!60!gray!10}0.142    & \cellcolor{cyan!60!gray!10}13.380     & \cellcolor{cyan!60!gray!10} 1.000  & \cellcolor{cyan!60!gray!10}	0.072  & \cellcolor{cyan!60!gray!10}10.770   \\ \hline  	

			\multirow{5}{*}{APIGraph}                       
			& RF                        & 1.000    & 0.039    & 11.260    & 1.000    & 0.098    & 14.930     & 1.000  &0.040  & 	9.530    \\  	 

			& \cellcolor{cyan!60!gray!10}DNN                       & \cellcolor{cyan!60!gray!10}1.000    & \cellcolor{cyan!60!gray!10}0.132    & \cellcolor{cyan!60!gray!10}14.370    & \cellcolor{cyan!60!gray!10}1.000    & \cellcolor{cyan!60!gray!10}0.096    & \cellcolor{cyan!60!gray!10}18.630   &  \cellcolor{cyan!60!gray!10}	1.000      &    \cellcolor{cyan!60!gray!10}0.168 
	     &  \cellcolor{cyan!60!gray!10}12.566 
      \\  	 

			& AB                        & 1.000    & 0.093    & 14.510    & 0.990    & 0.131    & 18.350     &  1.000  & 0.067   &	12.250   \\ 	 

			& \cellcolor{cyan!60!gray!10}1-NN                      & \cellcolor{cyan!60!gray!10}1.000   & \cellcolor{cyan!60!gray!10}0.058    & \cellcolor{cyan!60!gray!10}11.190    & \cellcolor{cyan!60!gray!10}1.000   & \cellcolor{cyan!60!gray!10}0.089    & \cellcolor{cyan!60!gray!10}14.040     & \cellcolor{cyan!60!gray!10}1.000  & \cellcolor{cyan!60!gray!10}0.012  & \cellcolor{cyan!60!gray!10}6.910    \\ 		

			& 3-NN                      & 1.000   & 0.085    & 11.570    & 1.000   & 0.105    & 13.770     & 1.000   & 	0.019 & 7.780   \\ \hline  	

			\multirow{1}{*}{GCN} & \cellcolor{cyan!60!gray!10}DNN                        & \cellcolor{cyan!60!gray!10}1.000   & \cellcolor{cyan!60!gray!10}0.205  & \cellcolor{cyan!60!gray!10}11.610   & \cellcolor{cyan!60!gray!10}1.000  & \cellcolor{cyan!60!gray!10}0.104   & \cellcolor{cyan!60!gray!10}17.320   & \cellcolor{cyan!60!gray!10}-  & \cellcolor{cyan!60!gray!10}-  & \cellcolor{cyan!60!gray!10}-   \\  \hline
			
		\end{tabular}%
	}
	\label{tab:effectiveness}
\end{table*}

\textbf{Target Model}.
We choose {three SOTA malware detection methods (i.e., MaMadroid \cite{DBLP:conf/ndss/MaricontiOACRS17}, APIGraph \cite{DBLP:conf/ccs/ZhangZZDCZZY20} and GCN \cite{DBLP:journals/corr/abs-2009-05602})} as our target system. 
In MaMadroid and APIGraph, we employ Random Forest (RF) \cite{Breiman2001Random}, AdaBoost (AB) \cite{2002Logistic}, 1-Nearest Neighbor (1-NN) \cite{1952Discriminatory}, 3-Nearest Neighbor (3-NN) and Dense Neural Network (DNN) as the target classifier, respectively. Similar to  \cite{DBLP:journals/corr/abs-2009-05602}, we use a two-layer DNN as the target classifier in the GCN-based method.

\textbf{Metric}.
We  use attack success rate (\textbf{ASR}), average perturbation ratio (\textbf{APR}), and the number of interaction rounds (\textbf{IR}) to evaluate {\ourtool}.
{
ASR corresponds to the ratio of the number of successfully generated AEs (denoted by $N_{success}$) to the number of malicious examples used for AE generation (denoted by $N_{total}$ ), i.e., $ASR=N_{success} / N_{total}$.
APR is the ratio of the number of added edges (denoted by $E_{added}$) to the total number of edges (denoted by $E_{total}$), i.e., $APR = E_{added} / E_{total}$. 
}
IR is defined as the number of interactions between our attack model and the target model.

\subsection{RQ1: Effectiveness}\label{sec:RQ1}

\textbf{Experimental Setup.} To verify the attack effectiveness of {\ourtool}, we use {\ourtool} to attack the {32} target models\footnote{{ Our experiments use 2 traditional 
FCG-based feature extraction methods (MaMaDroid and APIgraph), 3 feature levels (class, family and package) and 5 target classifiers (RF, AB, etc.). Furthermore, 1 GCN feature extraction method is considered with 2 feature levels (family and package). Hence, there are $32= 2\times 3\times5+1\times2 $ classifiers.}} mentioned above, and calculate ASR, APR and IR on every target model. 


{Furthermore, we also compare {\ourtool} with three attack methods, i.e., SRL \cite{DBLP:journals/corr/abs-2009-05602}, SRL\_N and Random Insertion (RI). To our knowledge, SRL is the SOTA malware AE generation method \footnote{{Note that SRL works on control flow graph instead of FCG. To apply SRL to the FCG based Android malware detection, we design a non-functional API list (instead of non-functional instruction list), which contains 17 non-functional APIs.}}.  Since SRL requires knowing the class probabilities outputted by the target model, we modify its reward function and create a variant of SRL (i.e., SRL\_N) that only relies on binary outputs. The RI attack method is also introduced from \cite{DBLP:journals/corr/abs-2009-05602}, and it randomly inserts non-functional functions. }

\begin{figure}[h]
	\centering
	\includegraphics[scale=0.7]{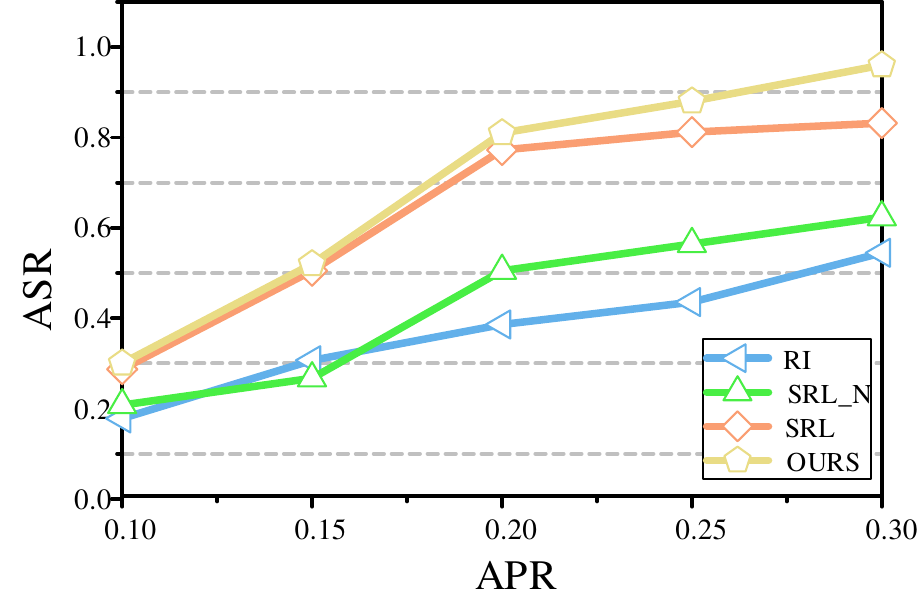}
	\caption{{Comparison with SOTA methods.}}
	\label{fig:SOTA}
\end{figure}

\textbf{Results \& Analyses}.
{Table \ref{tab:effectiveness} reflects the attack performance of {\ourtool} on MaMaDroid, APIGraph and GCN under various feature granularities.}
{First, {\ourtool} achieves an average ASR of 99.9\% over 32 target models, hence confirming the effectiveness of {\ourtool}.} { Second, when attacking the family-granularity classifier, {\ourtool} achieves the lowest APR and IR (i.e., 0.071 and 10.936).} {This indicates that although the family-granularity FCG speeds up malware detection through reducing input complexity, it still improves the efficiency of {\ourtool} by reducing search space.} 


Fig. \ref{fig:SOTA} compares {\ourtool}, SRL, SRL\_N and RI with respect to ASR under various APRs. Not surprisingly, RI performs worst in our experiments due to its poor search strategy. SRL performs better than SRL\_N, because SRL has access to class probabilities, which is more valuable than binary decisions. It is worth noting that {\ourtool} still outperforms SRL (e.g., its ASR is 4\% higher when APR is 0.2), although {\ourtool} cannot utilize class probabilities. 
 {The above results confirm that, under a certain number of perturbations, our method generates a better combination of the added edges that are more deceptive to detectors, as compared to the other methods.}



\subsection{RQ2: Evolution }
\textbf{Experimental Setup.}
In this subsection, we use experiments to analyze the effects of multi-population co-evolution mechanism.
First, we want to show that this mechanism can overcome the challenge of unknown feature granularity. To this end, we compare our method with the single-population methods in attacking MaMaDroid. The single-population methods rely on one single population corresponding to class, package and family  level, denoted by {\ourtool}-C, {\ourtool}-P  and {\ourtool}-F, respectively. {We also randomly select a malware sample and take a close look at these methods' attack processes.}

 Second, we want to know whether the correct feature granularity is found by our method.  We then record the survival number of each population and analyze how these populations evolve. { In this experiment, we choose MaMaDroid with an RF classifier as our target model, and use family-level feature granularity in malware detection. }


\begin{figure}[h]

	\centering
	\includegraphics[scale=0.55]{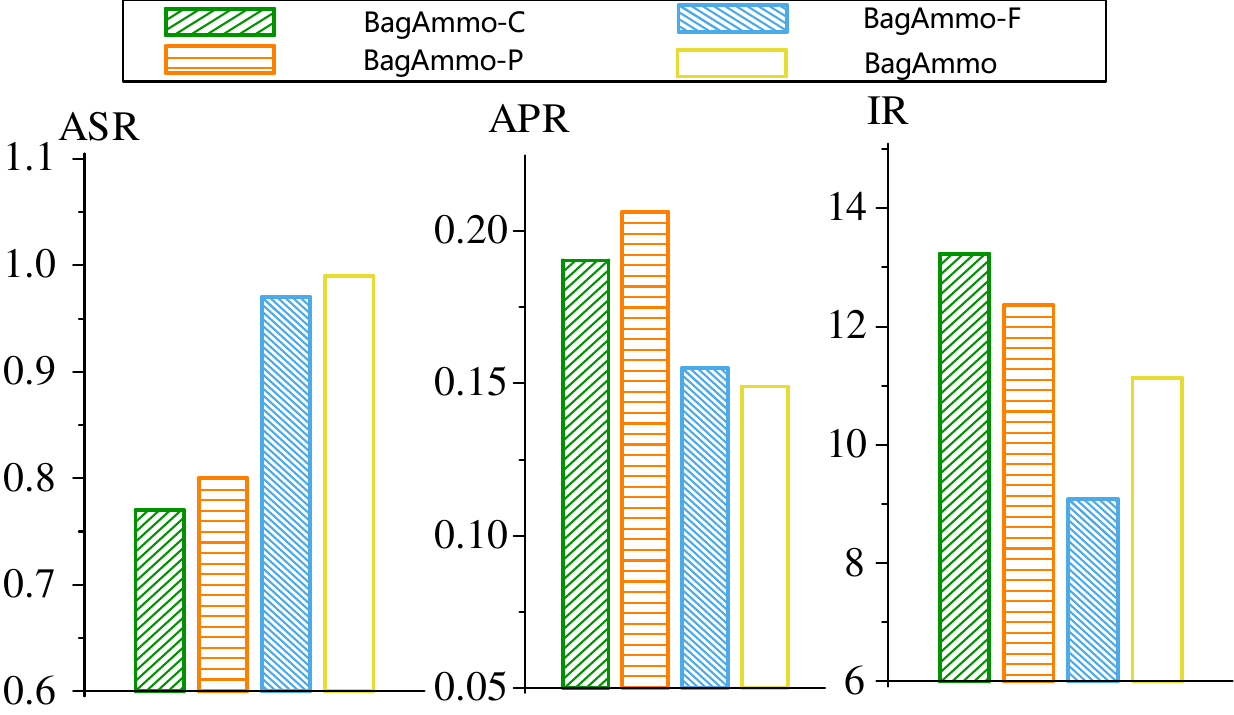}
	\caption{ {Performance comparison with single-population.}}
	\label{fig:multi}
\end{figure}

\begin{figure*}[h]
	\centering
	\includegraphics[scale=0.55]{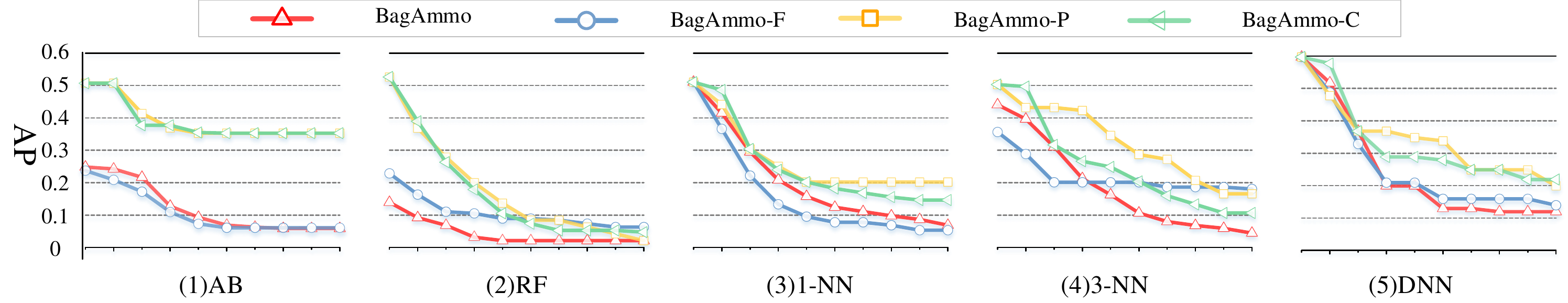}
	\caption{{Multi-population vs. single-population. }}
	\label{fig:exp_3}
\end{figure*}

\begin{figure}[h]
	\centering
	\includegraphics[scale=0.55]{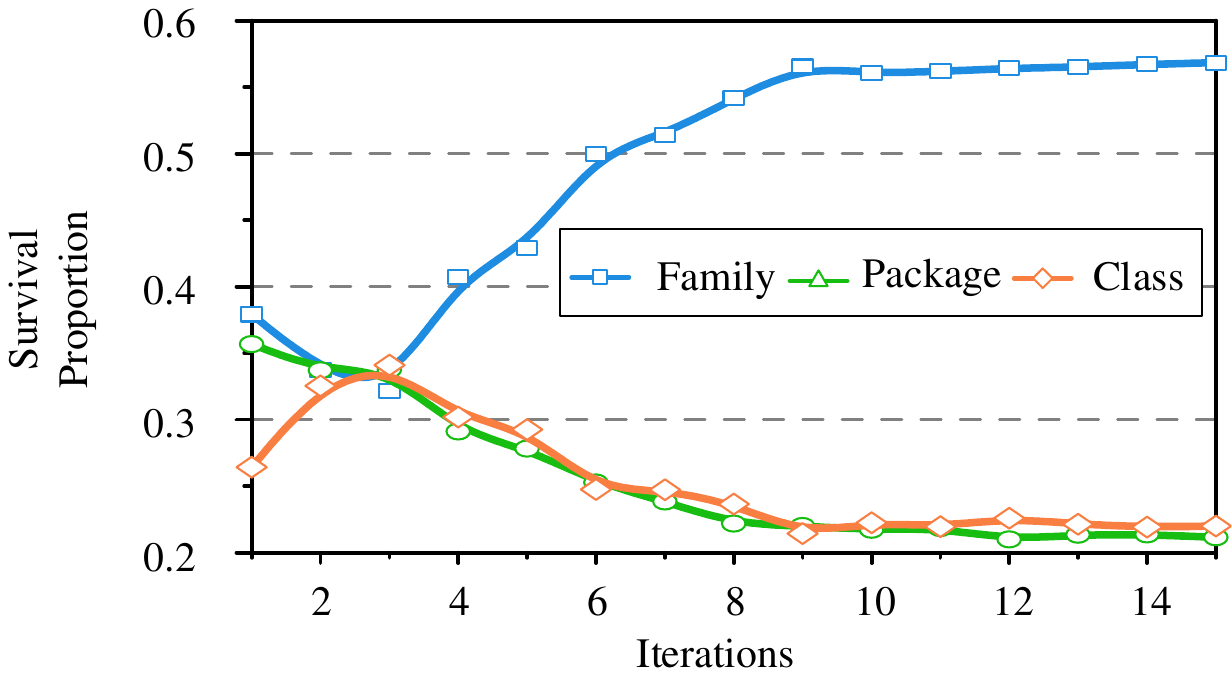}
	\caption{{The changing trend of survival proportion.}}
	\label{fig:living_number_change}
\end{figure}


\textbf{Results \& Analyses.} 
{ For comparison, we choose family-level feature granularity and evaluate {\ourtool} and single-population methods on \textbf{all}  test samples. The results are shown in Fig. \ref{fig:multi}. It can be seen that {\ourtool} performs best and achieves the highest ASR with the lowest APR. {\ourtool}-C and {\ourtool}-P perform worst since they use a false feature granularity. Surprisingly, {\ourtool} performs better than {\ourtool}-F (i.e., 2\% higher in ASR and 0.06 lower in APR). This is because the introduction of multiple populations helps to avoid premature convergence and approach a global optimum. However, it may result in more interactions with the target model. This accounts for why {\ourtool} has a higher IR than {\ourtool}-F.} 


{Now we randomly select a malware sample, and use it to generate an AE to attack 5 classifiers under family-level feature granularity. The attack processes of all methods are depicted in Fig. \ref{fig:exp_3}.} In this figure, the vertical axis represents the perturbation ratio of all methods, and the horizontal axis shows the IR values. {If a curve exhibits an evident decreasing trend and falls below a low threshold, we can conclude that the corresponding method succeeds in generating an AE and defeating the target model. As for those curves keeping horizontal (e.g., the green curve in the first subfigure), the corresponding methods fail to generate AEs. }
Fig.\ref{fig:exp_3} shows that the perturbation ratio of multi-population always has a satisfactory decreasing trend, hence confirming the effects of multi-population co-evolution.
Furthermore, using a single population may cause {premature convergence to a local optimum}, as indicated by Fig. \ref{fig:exp_3}-(1). However, {\ourtool} effectively mitigates this problem using  multiple populations. { Theoretical analyses are given in Appendix \ref{appendix: theoretical analyses}.}

Finally, we verify the multi-population  co-evolution method converges to the real feature granularity from a different perspective. We show the survival proportion (i.e., the ratio between the number of alive individuals and the total number of individuals) of different populations in Fig. \ref{fig:living_number_change}. At the beginning, the perturbations are randomly added, and the survival proportion of different populations are irregular. However, as the number of queries increases, the family population and class population gradually fall to a low level. Contrarily, the survival proportion of the population corresponding to the correct feature granularity (i.e., {family} level) gradually rises to a high level. This phenomenon also confirms the effects of multi-population co-evolution.

\subsection{RQ3: Efficiency}\label{sec:RQ2}
\textbf{Experimental Setup}. 
We conduct \textit{ablation studies} to verify the effects of the substitute model in decreasing queries and improving attack efficiency. For comparison, we remove the substitute model and guide the multi-population co-evolution algorithm only using the target model. This method is called {\ourtool}-Without-S. Then we use {\ourtool} and {\ourtool}-Without-S to manipulate the same APK file, {and compare their performance.} 

\begin{figure*}[h]
	\centering
	\includegraphics[scale=0.6]{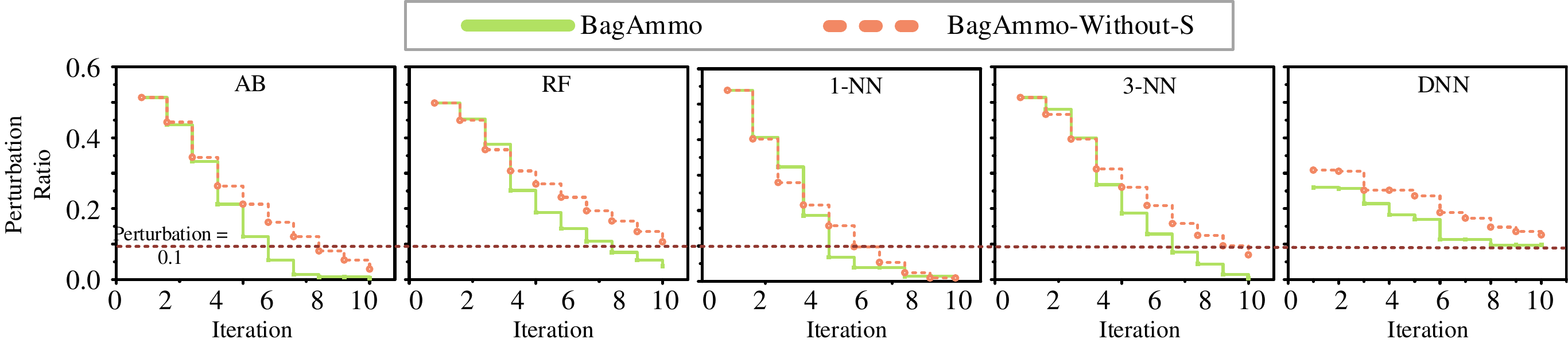}
	\caption{{With substitute model vs. without substitute model.}}
	\label{fig:with_and_without}
\end{figure*}


\begin{figure}[h]
	\centering
	\includegraphics[scale=0.55]{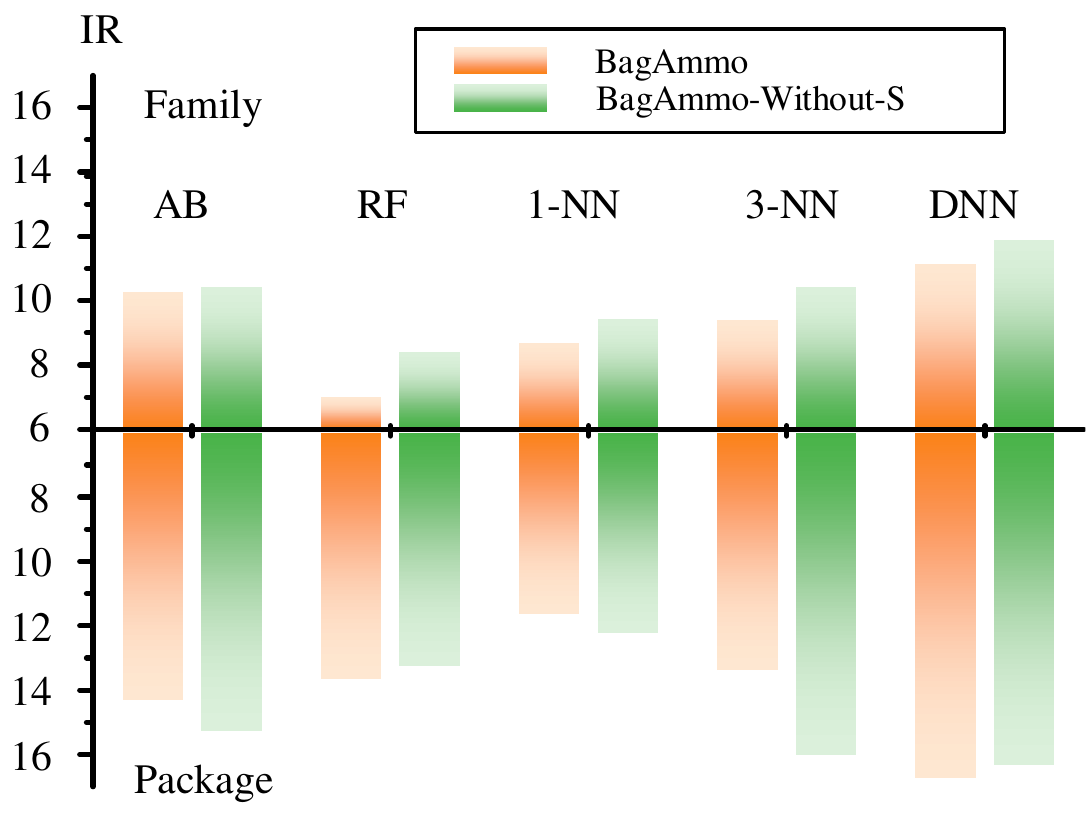}
	\caption{{Can substitute model reduce queries?}}
	\label{fig:exp_2_alldata}
\end{figure}


\textbf{Results \& Analyses}.
Substitute model's effects are shown in  Fig. \ref{fig:with_and_without}, where the solid and the dotted lines represent {\ourtool} and {\ourtool}-Without-S, respectively. {The vertical axis reflects perturbation ratio, and the horizontal axis indicates the number of queries}. It can be seen that in all cases, {\ourtool} always has a higher convergence speed. Moreover, {\ourtool} always requires fewer queries before the perturbation ratio is kept below a certain threshold (e.g., 0.1). 
Note that the difference between two methods in the initial phase is relatively small. It is because the substitute model has not been well trained in this phase. However, after the substitute model is well trained with sufficient data\footnote{ {In general,  the training accuracy of the substitute model arises as
the number of iteration rounds increase. However, the increasing trend of training accuracy is not strictly monotonic, because the training data used in the iterations are different. }}, {\ourtool} performs more efficiently and exhibits its advantage.

{Fig. \ref{fig:exp_2_alldata} compares {\ourtool } and {\ourtool }-Without-S in terms of IR. Its top-half part gives the results on the family-level FCG based MaMadroid, while the bottom-half part shows the results on the package-level FCG based MaMadroid. The horizontal axis indicates various classifiers (e.g., AB, RF and 1-NN). We can draw two conclusions from this figure. First, the package-level classifier is more difficult to attack. This is because package-level FCGs contain much more nodes than family-level FCGs, resulting in a larger search space for {\ourtool }. Second, using the substitute model reduces the number of queries in almost all cases and helps enhance  the attack efficiency.}

\subsection{RQ4:  {Manipulation Overhead}}
\textbf{Experimental Setup.}
Here we study the number of code modifications (i.e., manipulation overhead) required to generate a real evasive malware. We use experimental results to reflect the relationship between ASR and {the allowed perturbation ratio}. 

\textbf{Result \& Analysis.}
Experimental results are shown in Fig. \ref{fig:perturbation}. In this figure, the horizontal axis represents the allowed perturbation ratio, and the vertical axis gives the cumulative distribution function (CDF) of ASR. It can be observed that the ASR keeps rising with the increase of the allowed perturbation ratio. In practice, a larger perturbation ratio results in larger computational overhead for adversaries. Therefore, there exists a trade-off between manipulation overhead and attack success ratio. Moreover, 
 the 3-NN classifier is more robust than the 1-NN classifier. {It is because} the 3-NN classifier considers more data than 1-NN classifier when classifying a sample, which makes it distinguish benign and malicious apps easier. 


\begin{figure}[h]
	\centering
	\includegraphics[scale=0.4]{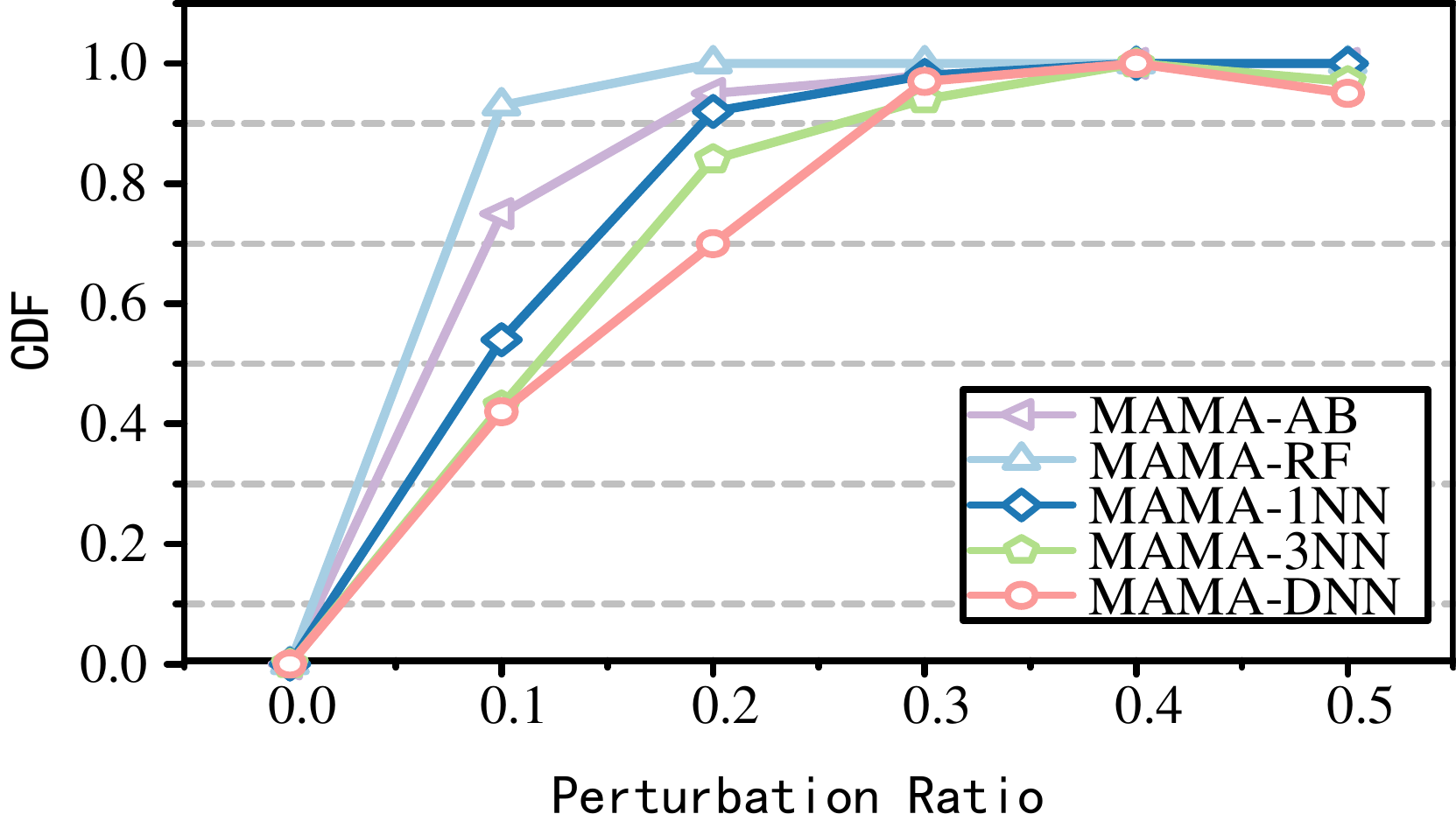}
	\caption{{CDF of ASR under different perturbation ratios.}}
	\label{fig:perturbation}
\end{figure}


\subsection{RQ5: Resilience}
\label{subsection: resilience}
\textbf{Experimental Setup.}
Concept drift \cite{263854} is often observed in the realistic applications of Android malware detection. 
 {Concept drift undermines existing AE generation methods that insert APIs selected from a pre-determined white list, if the white list is not updated accordingly. Hence we want to know whether {\ourtool} is also susceptible to concept drift}. To this end, we use newly-emerged malware samples to generate AEs and attack the classifiers trained over old data. We divide the dataset into training sets and testing sets according to the years that Android app emerge. We construct four new datasets to evaluate {\ourtool} under concept drift, as depicted in Table \ref{tab:setting}. {The first row of Table \ref{tab:setting} is the year of training samples used for training target classifiers. The second row is the year of testing samples used for generating AEs. The third row is the accuracy of the classifier.}

{Data imbalance is another practical problem worthy of consideration\cite{277204,DBLP:conf/uss/PendleburyPJKC19}. Since malicious samples are more difficult to collect than benign samples, malware detection models are usually trained over imbalanced data. We want to know whether data imbalance negatively impacts the attack performance of {\ourtool}. Hence we evaluate {\ourtool} on the target model trained with imbalanced data (the benign-malicious ratio is 10:1). }

\begin{table}
	\caption{{The attack performance under concept drift.}}
	\renewcommand\arraystretch{1.5}
	\centering
	\scalebox{0.7}{
		\begin{tabular}{ccccc}
			\hline		
			Training set (year) & \cellcolor{cyan!60!gray!10}2016    & \cellcolor{cyan!60!gray!10}2016-2017 & \cellcolor{cyan!60!gray!10}2016-2018 & \cellcolor{cyan!60!gray!10}2016-2019 \\ 
			Testing set (year)  & 2017    & 2018      & 2019      & 2020      \\ 
			ACC          & 92.92\% & 94.08\% & 94.58\% & 95.47\% \\  \hline
			ASR          & \cellcolor{cyan!60!gray!10}100\%   & \cellcolor{cyan!60!gray!10}100\%   & \cellcolor{cyan!60!gray!10}100\%   & \cellcolor{cyan!60!gray!10}100\%  \\ \hline
		\end{tabular}%
		\label{tab:setting}
	}
\end{table}

\begin{table}[]
	\caption{{The APR on balanced and imbalanced data.}}
	\renewcommand\arraystretch{1.5}
	\centering
	\scalebox{0.7}{
\begin{tabular}{ccccccc}
\hline
Level & Case & AB & RF & 1NN & 3NN & Average \\ \hline
\multirow{2}{*}{Family} & \cellcolor{cyan!60!gray!10}Balance & \cellcolor{cyan!60!gray!10}0.066 & \cellcolor{cyan!60!gray!10}0.021 & \cellcolor{cyan!60!gray!10}0.031 & \cellcolor{cyan!60!gray!10}0.037 & \cellcolor{cyan!60!gray!10}0.039 \\
 & Imbalance & 0.050 & 0.024 & 0.021 & 0.021 & 0.029 \\ \hline
\multirow{2}{*}{Package} & \cellcolor{cyan!60!gray!10}Balance & \cellcolor{cyan!60!gray!10}0.072 & \cellcolor{cyan!60!gray!10}0.049 & \cellcolor{cyan!60!gray!10}0.109 & \cellcolor{cyan!60!gray!10}0.142 & \cellcolor{cyan!60!gray!10}0.093 \\
 & Imbalance & 0.041 & 0.040 & 0.094 & 0.105 & 0.070 \\ \hline
\end{tabular}%
}
\label{fig:imbalance}
\end{table}


\textbf{Result \& Analysis.}
{In the experiments on concept drift, we use {\ourtool} to manipulate the test samples, and the results are used to attack the MaMaDroid model with the family-level feature and the classifier of RF}. The ASR of {\ourtool} in every scenario is presented in the last row of Table \ref{tab:setting}. First, this table indicates that with more training samples, the accuracy of the target classifier becomes higher. No matter how high the accuracy is, however, {\ourtool} always achieves a perfect ASR of 100\%. 
 {This shows that {\ourtool} performs well under concept drift and is still efficient when the malware detection models learn more with the new data.}
Note that {\ourtool} uses the functions coming from the malware itself (instead of a static function set).  {\ourtool} reduces the risk of using functions that become outdated due to concept drift. As a result, {\ourtool} poses a persistent threat to malware detectors. 
     {Finally, we also discuss how {\ourtool} performs when the defender has the knowledge of adversarial example in Appendix \ref{app:ar}}

{Table \ref{fig:imbalance} shows the experimental results of {\ourtool} in the cases of balanced and imbalanced data. Our experiments demonstrate that the DNN model performs very poorly when trained with imbalanced data. Therefore, we do not choose DNN as our target model. 
 {In both cases (i.e., balanced dataset and imbalance dataset), {\ourtool} achieves an attack success rate (i.e., ASR) of 100\%. Here we only show the values of APR in Table \ref{fig:imbalance}.}
A higher APR means a more difficult attack task. It can be seen that in the vast majority of cases, {\ourtool} needs fewer perturbations (i.e., has a lower APR) to attack the target model trained with imbalanced data. That is, data imbalance does not bring troubles to {\ourtool}. This is because that training with imbalanced data makes the target model more likely to classify malware as benign apps. Accordingly, this reduces the degree of difficulty in generating AEs.



\subsection{{RQ6: Functionality} }
{\textbf{Experimental Setup.}
In this section, we first use static analysis to verify whether the perturbations generated by {\ourtool} are successfully imposed on malware. We then employ dynamic analysis to check whether the perturbation changes the functionality of the malware.}

{\textbf{Result \& Analysis.}
To know whether our perturbations are injected, we add a unique log statement when a perturbation (i.e., a try-catch trap) is injected. This log statement helps us to find the perturbation in the smali file. We then check if the found function calls in the smali file coincide with the perturbations generated by {\ourtool}. In our experiments, we evaluate 50 APK files, and we realize that all the generated perturbations are correctly injected into the smali file.
}

{ In our experiments of dynamic analysis, we first install and run 50 pairs of original and perturbed malware samples in Android Virtual Device (AVD). It is observed that every malware pair performs the same and has the same run-time UI. For further analysis, we insert three log statements, denoted by LOG1, LOG2 and LOG3,  into every try-catch block to record execution information. LOG1 is in front of the runtime exception, LOG2 is in front of the inserted function, and LOG3 is at the beginning of the catch block. We analyze 50 APK files aided by  Android Studio's log analysis tool (i.e., LogCat). We realize that either LOG1 or LOG3 of every APK file is normally executed, but no LOG2 is executed. This phenomenon means that all manipulated malware samples run properly, and the inserted functions are not invoked, hence posing no impact on the malware functionality.
}


	\section{Related Work}
\label{sec:Related}


{ Recently, adversarial attacks have been widely used in various fields, i.e., image classification \cite{ijcai2022p554,xia2022enhancing}, traffic analysis \cite{DBLP:conf/uss/NasrBH21,DBLP:journals/tifs/RahmanIMW21}, autonomous driving \cite{DBLP:conf/uss/Jing0D0L0NW21,DBLP:journals/corr/abs-2201-06192} and object detection \cite{DBLP:conf/uss/LovisottoTSSM21}. As for Android malware detection,}
there have been many studies \cite{huang2018adversarial,grosse2017adversarial,hu2017generating,2019lh,DBLP:journals/tifs/LiL20} on syntax features oriented AE generation. Huang \emph{et al.} \cite{huang2018adversarial}  use the saddle-point optimization formulation to generate adversarial examples in the discrete (e.g., binary) domain for malware detection. Grosse  \emph{et al.} \cite{grosse2017adversarial}  expand existing AE generation algorithms to construct a highly effective attack against malware detection models. In \cite{hu2017generating,2019lh}, Hu \emph{et al.} utilize a GAN to generate adversarial examples in black-box mode for malware detection. Li  \emph{et al.} \cite{ DBLP:journals/tifs/LiL20} propose { an ensemble approach that allows} attackers to perturb a malware example via multiple attack methods and multiple manipulation sets.

To achieve higher detection accuracy, more and more {Android malware detection methods} \cite{DBLP:conf/ccs/ZhangZZDCZZY20,DBLP:conf/ndss/MaricontiOACRS17,DBLP:conf/kbse/WuLZYZ019} focus on semantic features. Chen  \emph{et al.} \cite{chen2020android} introduce two AE generations methods in image classification to detect Android malware, and propose a method applying optimal perturbations onto Android APKs. Their method directly perturbs features in feature space. Pierazzi  \emph{et al.} \cite{DBLP:conf/sp/PierazziPCC20}
extract slices of bytecode (i.e., gadgets) from benign APKs and inject them into a malicious APK to generate adversarial malware. { Zhang \emph{et al.} \cite{DBLP:journals/corr/abs-2009-05602} propose a reinforcement learning based attack to deceive graph feature based malware detection models.  Recently, Bostani \emph{et al.} \cite{DBLP:journals/corr/abs-2110-03301} propose an interesting black-box attack EvadeDroid without requiring the knowledge about feature space. Different from {\ourtool}, EvadeDroid employs random search to find the desired perturbation from the code of benign apps.}

\section{ {Limitations and Discussion}}
\label{sec:Limitations}
 {In this paper, we propose a black-box AE attack BagAmmo towards the FCG based Android malware detection. We hope that our work has reference value for the
study of Android malware detection, and raises the concern for the threats posed by AE attacks. Moreover, our method can be used to evaluate the robustness of existing Android malware detection
methods. Below we discuss some limitations and future works.}

 {\textbf{Dynamic analysis based defense}. 
Our method targets  the static analyse methods.
 It relies on inserting function calls to change FCG. But it does not change the information flow of malware. Therefore, it does not negatively impact dynamic analysis \cite{2017Malton}. We will explore how to construct adversarial examples against dynamic analysis based Android malware detection methods in future work.}

 
\textbf{Transfer to other domains}. The idea and the framework of {\ourtool} are transferable to a certain degree, since many domains  use semantic features and graph structured data (e.g., intrusion detection system \cite{DBLP:journals/iotj/ZhouLLYSW22} and trajectory prediction system \cite{wong2022view,xia2022cscnet}).

  {\textbf{Try/catch detection based defense}. Another concern is that whether a defender can detect the AEs generated by {\ourtool} by counting the number of try/catch blocks. This defense method requires a detection threshold for the number of try-catch blocks. Through comparing the try-catch block number of original malicious APKs and that of adversarially perturbed APKs, however, we find that the number of try-catch blocks added by our method is relatively small. So it is difficult to find an appropriate threshold for all APKs. Without such a threshold, this defense method may cause a high false positive or false negative rate.\footnote{ {More experimental results can be found in Appendix \ref{app:trycatch}}}} 

  \section{ Acknowledgments
}
This work was supported partially by the  Hong Kong RGC Project  (No.  PolyU15219319), HKPolyU Grant No.ZVG0,   Fundamental Research Funds for the Central Universities (HUST: Grant No. YCJJ202202016 and  2022JYCXJJ035) .

\bibliographystyle{unsrt}

\bibliography{ref}{}

\begin{thebibliography}{10}

\bibitem{DBLP:conf/eccv/AndriushchenkoC20}
Maksym Andriushchenko, Francesco Croce, Nicolas Flammarion, and Matthias Hein.
\newblock Square attack: {A} query-efficient black-box adversarial attack via
  random search.
\newblock In {\em Proc. {ECCV}}, 2020.

\bibitem{277204}
Daniel Arp, Erwin Quiring, Feargus Pendlebury, Alexander Warnecke, Fabio
  Pierazzi, Christian Wressnegger, Lorenzo Cavallaro, and Konrad Rieck.
\newblock Dos and don'ts of machine learning in computer security.
\newblock In {\em Proc. { } Security}, 2022.

\bibitem{DBLP:conf/ndss/ArpSHGR14}
Daniel Arp, Michael Spreitzenbarth, Malte Hubner, Hugo Gascon, and Konrad
  Rieck.
\newblock {DREBIN:} effective and explainable detection of android malware in
  your pocket.
\newblock In {\em Proc. {NDSS}}, 2014.

\bibitem{DBLP:conf/icse/BaiX00M20}
Yude Bai, Zhenchang Xing, Xiaohong Li, Zhiyong Feng, and Duoyuan Ma.
\newblock Unsuccessful story about few shot malware family classification and
  siamese network to the rescue.
\newblock In {\em Proc. {ICSE}}, 2020.

\bibitem{DBLP:conf/icml/BojchevskiG19}
Aleksandar Bojchevski and Stephan G{\"{u}}nnemann.
\newblock Adversarial attacks on node embeddings via graph poisoning.
\newblock In {\em Proc. {ICML}}, 2019.

\bibitem{DBLP:journals/corr/abs-2110-03301}
Hamid Bostani and Veelasha Moonsamy.
\newblock Evadedroid: {A} practical evasion attack on machine learning for
  black-box android malware detection.
\newblock {\em CoRR}, abs/2110.03301, 2021.

\bibitem{Breiman2001Random}
Breiman.
\newblock Random forests.
\newblock {\em MACH LEARN}, 2001,45(1)(-):5--32, 2001.

\bibitem{DBLP:journals/ijon/CaiJGLY21}
Minghui Cai, Yuan Jiang, Cuiying Gao, Heng Li, and Wei Yuan.
\newblock Learning features from enhanced function call graphs for android
  malware detection.
\newblock {\em Neurocomputing}, 423:301--307, 2021.

\bibitem{DBLP:conf/sp/Carlini017}
Nicholas Carlini and David~A. Wagner.
\newblock Towards evaluating the robustness of neural networks.
\newblock In {\em Proc. {S\&P}}, 2017.

\bibitem{chen2020android}
Xiao Chen, Chaoran Li, Derui Wang, Sheng Wen, Jun Zhang, Surya Nepal, Yang
  Xiang, and Kui Ren.
\newblock Android hiv: A study of repackaging malware for evading
  machine-learning detection.
\newblock {\em {IEEE} Trans. Inf. Forensics Secur.}, 15:987--1001, 2020.

\bibitem{2002Logistic}
M.~Collins, R.~E. Schapire, and Y.~Singer.
\newblock Logistic regression, adaboost and bregman distances.
\newblock {\em Machine Learning}, 48(1/2/3):253--285, 2002.

\bibitem{DBLP:conf/icml/DaiLTHWZS18}
Hanjun Dai, Hui Li, Tian Tian, Xin Huang, Lin Wang, Jun Zhu, and Le~Song.
\newblock Adversarial attack on graph structured data.
\newblock In {\em Proc. {ICML}}, 2018.

\bibitem{DBLP:journals/tdsc/DemontisMBMARCG19}
Ambra Demontis, Marco Melis, Battista Biggio, Davide Maiorca, Daniel Arp,
  Konrad Rieck, Igino Corona, Giorgio Giacinto, and Fabio Roli.
\newblock Yes, machine learning can be more secure! {A} case study on android
  malware detection.
\newblock {\em {IEEE} Trans. Dependable Secur. Comput.}, 16(4):711--724, 2019.

\bibitem{DBLP:conf/ccs/EnckOM09}
William Enck, Machigar Ongtang, and Patrick~D. McDaniel.
\newblock On lightweight mobile phone application certification.
\newblock In {\em Proc. {CCS}}, pages 235--245, 2009.

\bibitem{DBLP:journals/tifs/Fan0LCTZL18}
Ming Fan, Jun Liu, Xiapu Luo, Kai Chen, Zhenzhou Tian, Qinghua Zheng, and Ting
  Liu.
\newblock Android malware familial classification and representative sample
  selection via frequent subgraph analysis.
\newblock {\em {IEEE} Trans. Inf. Forensics Secur.}, 13(8):1890--1905, 2018.

\bibitem{DBLP:conf/uss/PendleburyPJKC19}
{Feargus Pendlebury and Fabio Pierazzi and Roberto Jordaney and Johannes Kinder
  and Lorenzo Cavallaro}.
\newblock {{TESSERACT:} Eliminating Experimental Bias in Malware Classification
  across Space and Time}.
\newblock In {\em {Proc. {USENIX} Security}}, {2019}.

\bibitem{DBLP:conf/ndss/FengBMDA17}
Yu~Feng, Osbert Bastani, Ruben Martins, Isil Dillig, and Saswat Anand.
\newblock Automated synthesis of semantic malware signatures using maximum
  satisfiability.
\newblock In {\em Proc. {NDSS}}, 2017.

\bibitem{DBLP:conf/ntms/FereidooniCYS16}
Hossein Fereidooni, Mauro Conti, Danfeng Yao, and Alessandro Sperduti.
\newblock {ANASTASIA:} android malware detection using static analysis of
  applications.
\newblock In {\em Proc. {NTMS}}, 2016.

\bibitem{1952Discriminatory}
Evelyn Fix and J.~L Hodges, Jr.
\newblock Discriminatory analysis - nonparametric discrimination: Small sample
  performance.
\newblock 1952.

\bibitem{gdata}
Gata.
\newblock Mobile malware report - no let-up with android malware.
\newblock \url{https://www.gdatasoftware.com}, 2019.
\newblock Accessed April 4, 2010.

\bibitem{grosse2017adversarial}
Kathrin Grosse, Nicolas Papernot, Praveen Manoharan, M~Backes, and Patrick
  Mcdaniel.
\newblock Adversarial examples for malware detection.
\newblock In {\em Proc. ESORICS}, 2017.

\bibitem{guardsquare}
guardsquare.
\newblock The industry-leading java optimizer for android apps.
\newblock \url{https://www.guardsquare.com/proguard}.

\bibitem{DBLP:conf/kdd/HouYSA17}
Shifu Hou, Yanfang Ye, Yangqiu Song, and Melih Abdulhayoglu.
\newblock Hindroid: An intelligent android malware detection system based on
  structured heterogeneous information network.
\newblock In {\em Proc. {SIGKDD}}, 2017.

\bibitem{hu2017generating}
Weiwei Hu and Ying Tan.
\newblock Generating adversarial malware examples for black-box attacks based
  on gan.
\newblock {\em arXiv: 1702.05983}, 2017.

\bibitem{DBLP:conf/icccn/HuTMZZH14}
Wenjun Hu, Jing Tao, Xiaobo Ma, Wenyu Zhou, Shuang Zhao, and Ting Han.
\newblock Migdroid: Detecting app-repackaging android malware via method
  invocation graph.
\newblock In {\em Proc. {ICCCN}}, 2014.

\bibitem{HU2022108824}
Zichao Hu, Heng Li, Liheng Yuan, Zhang Cheng, Wei Yuan, and Ming Zhu.
\newblock Model scheduling and sample selection for ensemble adversarial
  example attacks.
\newblock {\em Pattern Recognition}, 130:108824, 2022.

\bibitem{huang2018adversarial}
Alex Huang, Abdullah Aldujaili, Erik Hemberg, and Unamay Oreilly.
\newblock Adversarial deep learning for robust detection of binary encoded
  malware.
\newblock In {\em Proc. IEEE S\&P Workshops}, 2018.

\bibitem{DBLP:journals/corr/abs-2201-06192}
Wei Jia, Zhaojun Lu, Haichun Zhang, Zhenglin Liu, Jie Wang, and Gang Qu.
\newblock Fooling the eyes of autonomous vehicles: Robust physical adversarial
  examples against traffic sign recognition systems.
\newblock {\em CoRR}, abs/2201.06192, 2022.

\bibitem{DBLP:conf/uss/Jing0D0L0NW21}
Pengfei Jing, Qiyi Tang, Yuefeng Du, Lei Xue, Xiapu Luo, Ting Wang, Sen Nie,
  and Shi Wu.
\newblock Too good to be safe: Tricking lane detection in autonomous driving
  with crafted perturbations.
\newblock In {\em Proc. {USENIX} Security}, 2021.

\bibitem{DBLP:conf/icssa/KimL18}
Hyung{-}Jong Kim and Hae~Young Lee.
\newblock A study on the privacy protection layer for android iot services
  (lightning talk).
\newblock In {\em Proc. ICSSA}, 2018.

\bibitem{2017Malton}
X.~Lei, Y.~Zhou, T.~Chen, X.~Luo, and G.~Gu.
\newblock Malton: Towards on-device non-invasive mobile malware analysis for
  art.
\newblock In {\em Proc. USENIX Security}, 2017.

\bibitem{1997Degree}
Y.~Leung and G.~Yong.
\newblock Degree of population diversity - a perspective on premature
  convergence in genetic algorithms and its markov chain analysis.
\newblock {\em IEEE Transactions on Neural Networks}, 8(5):1165--1176, 1997.

\bibitem{DBLP:journals/tifs/LiL20}
Deqiang Li and Qianmu Li.
\newblock Adversarial deep ensemble: Evasion attacks and defenses for malware
  detection.
\newblock {\em {IEEE} Trans. Inf. Forensics Secur.}, 15:3886--3900, 2020.

\bibitem{2019lh}
Heng Li, ShiYao Zhou, Wei Yuan, and Henry Leung.
\newblock Adversarial-example attacks towards android malware detection system.
\newblock {\em IEEE Systems Journal}, 2019.

\bibitem{DBLP:journals/tii/LiSYLSY18}
Jin Li, Lichao Sun, Qiben Yan, Zhiqiang Li, Witawas Srisa{-}an, and Heng Ye.
\newblock Significant permission identification for machine-learning-based
  android malware detection.
\newblock {\em {IEEE} Trans. Ind. Informatics}, 14(7):3216--3225, 2018.

\bibitem{DBLP:conf/icdm/LiYZ18}
Pengcheng Li, Jinfeng Yi, and Lijun Zhang.
\newblock Query-efficient black-box attack by active learning.
\newblock In {\em Proc. {ICDM}}, 2018.

\bibitem{DBLP:journals/tmc/LiuLZWZ20}
Xing Liu, Jiqiang Liu, Sencun Zhu, Wei Wang, and Xiangliang Zhang.
\newblock Privacy risk analysis and mitigation of analytics libraries in the
  android ecosystem.
\newblock {\em {IEEE} Trans. Mob. Comput.}, 19(5):1184--1199, 2020.

\bibitem{DBLP:conf/iclr/LiuCLS17}
Yanpei Liu, Xinyun Chen, Chang Liu, and Dawn Song.
\newblock Delving into transferable adversarial examples and black-box attacks.
\newblock In {\em Proc. {ICLR}}, 2017.

\bibitem{DBLP:conf/uss/LovisottoTSSM21}
Giulio Lovisotto, Henry Turner, Ivo Sluganovic, Martin Strohmeier, and Ivan
  Martinovic.
\newblock {SLAP:} improving physical adversarial examples with short-lived
  adversarial perturbations.
\newblock In {\em Proc. {USENIX} Security}, 2021.

\bibitem{DBLP:conf/kdd/0001WDWT21}
Yao Ma, Suhang Wang, Tyler Derr, Lingfei Wu, and Jiliang Tang.
\newblock Graph adversarial attack via rewiring.
\newblock In {\em Proc. {KDD}}, 2021.

\bibitem{DBLP:conf/ndss/MaricontiOACRS17}
Enrico Mariconti, Lucky Onwuzurike, Panagiotis Andriotis, Emiliano~De
  Cristofaro, Gordon~J. Ross, and Gianluca Stringhini.
\newblock Mamadroid: Detecting android malware by building markov chains of
  behavioral models.
\newblock In {\em Proc. {NDSS}}, 2017.

\bibitem{DBLP:conf/acsac/MoserKK07}
Andreas Moser, Christopher Kruegel, and Engin Kirda.
\newblock Limits of static analysis for malware detection.
\newblock In {\em Proc. ACSAC}, 2007.

\bibitem{DBLP:conf/uss/NasrBH21}
Milad Nasr, Alireza Bahramali, and Amir Houmansadr.
\newblock Defeating dnn-based traffic analysis systems in real-time with blind
  adversarial perturbations.
\newblock In {\em Proc. {USENIX} Security}, 2021.

\bibitem{DBLP:conf/uss/OcteauMJBBKT13}
Damien Octeau, Patrick~D. McDaniel, Somesh Jha, Alexandre Bartel, Eric Bodden,
  Jacques Klein, and Yves~Le Traon.
\newblock Effective inter-component communication mapping in android: An
  essential step towards holistic security analysis.
\newblock In {\em Proc. {USENIX} Security}, 2013.

\bibitem{DBLP:conf/sp/PierazziPCC20}
Fabio Pierazzi, Feargus Pendlebury, Jacopo Cortellazzi, and Lorenzo Cavallaro.
\newblock Intriguing properties of adversarial {ML} attacks in the problem
  space.
\newblock In {\em Proc. {S\&P}}, 2020.

\bibitem{DBLP:journals/tifs/RahmanIMW21}
Mohammad~Saidur Rahman, Mohsen Imani, Nate Mathews, and Matthew Wright.
\newblock Mockingbird: Defending against deep-learning-based website
  fingerprinting attacks with adversarial traces.
\newblock {\em {IEEE} Trans. Inf. Forensics Secur.}, 16:1594--1609, 2021.

\bibitem{DBLP:journals/jnca/Seo0SBY14}
Seung{-}Hyun Seo, Aditi Gupta, Asmaa~Mohamed Sallam, Elisa Bertino, and Kangbin
  Yim.
\newblock Detecting mobile malware threats to homeland security through static
  analysis.
\newblock {\em J. Netw. Comput. Appl.}, 38:43--53, 2014.

\bibitem{DBLP:conf/www/SunWTHH20}
Yiwei Sun, Suhang Wang, Xianfeng Tang, Tsung{-}Yu Hsieh, and Vasant~G. Honavar.
\newblock Adversarial attacks on graph neural networks via node injections: {A}
  hierarchical reinforcement learning approach.
\newblock In {\em Proc. {WWW}}, 2020.

\bibitem{DBLP:conf/uss/SunSLM21}
Zhichuang Sun, Ruimin Sun, Long Lu, and Alan Mislove.
\newblock Mind your weight(s): {A} large-scale study on insufficient machine
  learning model protection in mobile apps.
\newblock In {\em Proc. {USENIX} Security}, 2021.

\bibitem{DBLP:conf/uss/SuyaC0020}
Fnu Suya, Jianfeng Chi, David Evans, and Yuan Tian.
\newblock Hybrid batch attacks: Finding black-box adversarial examples with
  limited queries.
\newblock In {\em Proc. {USENIX} Security}, 2020.

\bibitem{DBLP:journals/corr/SzegedyZSBEGF13}
Christian Szegedy, Wojciech Zaremba, Ilya Sutskever, Joan Bruna, Dumitru Erhan,
  Ian~J. Goodfellow, and Rob Fergus.
\newblock Intriguing properties of neural networks.
\newblock In {\em Proc. {ICLR}}, 2014.

\bibitem{DBLP:journals/corr/TanayG16}
Thomas Tanay and Lewis~D. Griffin.
\newblock A boundary tilting persepective on the phenomenon of adversarial
  examples.
\newblock {\em CoRR}, abs/1608.07690, 2016.

\bibitem{2012Genetic}
Anita Thengade and Rucha Dondal.
\newblock Genetic algorithm – survey paper.
\newblock {\em Foundation of Computer Science (FCS)}, 2012.

\bibitem{VirusTotal}
VirusTotal.
\newblock Virustotal - free online virus, malware and url scanner.

\bibitem{DBLP:journals/tifs/WangWFLHZ14}
Wei Wang, Xing Wang, Dawei Feng, Jiqiang Liu, Zhen Han, and Xiangliang Zhang.
\newblock Exploring permission-induced risk in android applications for
  malicious application detection.
\newblock {\em {IEEE} Trans. Inf. Forensics Secur.}, 9(11):1869--1882, 2014.

\bibitem{wong2022view}
Conghao Wong, Beihao Xia, Ziming Hong, Qinmu Peng, Wei Yuan, Qiong Cao, Yibo
  Yang, and Xinge You.
\newblock View vertically: A hierarchical network for trajectory prediction via
  fourier spectrums.
\newblock In {\em Proc. ECCV}, 2022.

\bibitem{DBLP:conf/ijcai/Wu0TDLZ19}
Huijun Wu, Chen Wang, Yuriy Tyshetskiy, Andrew Docherty, Kai Lu, and Liming
  Zhu.
\newblock Adversarial examples for graph data: Deep insights into attack and
  defense.
\newblock In {\em Proc. {IJCAI}}, 2019.

\bibitem{DBLP:conf/kbse/WuLZYZ019}
Yueming Wu, Xiaodi Li, Deqing Zou, Wei Yang, Xin Zhang, and Hai Jin.
\newblock Malscan: Fast market-wide mobile malware scanning by social-network
  centrality analysis.
\newblock In {\em Proc. {ASE}}, pages 139--150, 2019.

\bibitem{xia2022cscnet}
Beihao Xia, Conghao Wong, Qinmu Peng, Wei Yuan, and Xinge You.
\newblock Cscnet: Contextual semantic consistency network for trajectory
  prediction in crowded spaces.
\newblock {\em Pattern Recognition}, 126:108552, 2022.

\bibitem{ijcai2022p554}
Pengfei Xia, Ziqiang Li, Wei Zhang, and Bin Li.
\newblock Data-efficient backdoor attacks.
\newblock In {\em Proc. IJCAI}, 2022.

\bibitem{xia2022enhancing}
Pengfei Xia, Hongjing Niu, Ziqiang Li, and Bin Li.
\newblock Enhancing backdoor attacks with multi-level mmd regularization.
\newblock {\em IEEE Trans. Dependable Secur. Comput.}, 2022.

\bibitem{DBLP:conf/ijcai/XuC0CWHL19}
Kaidi Xu, Hongge Chen, Sijia Liu, Pin{-}Yu Chen, Tsui{-}Wei Weng, Mingyi Hong,
  and Xue Lin.
\newblock Topology attack and defense for graph neural networks: An
  optimization perspective.
\newblock In {\em Proc. {IJCAI}}, 2019.

\bibitem{263854}
Limin Yang, Wenbo Guo, Qingying Hao, Arridhana Ciptadi, Ali Ahmadzadeh, Xinyu
  Xing, and Gang Wang.
\newblock {CADE:} detecting and explaining concept drift samples for security
  applications.
\newblock In {\em Proc. {USENIX} Security}, 2021.

\bibitem{DBLP:journals/tsmc/YuanJLC21}
Wei Yuan, Yuan Jiang, Heng Li, and Minghui Cai.
\newblock A lightweight on-device detection method for android malware.
\newblock {\em {IEEE} Trans. Syst. Man Cybern. Syst.}, 51(9):5600--5611, 2021.

\bibitem{DBLP:journals/corr/abs-2009-05602}
Lan Zhang, Peng Liu, Yoonho Choi, and Ping Chen.
\newblock Semantics-preserving reinforcement learning attack against graph
  neural networks for malware detection.
\newblock {\em IEEE Trans Dependable Secure Comput}, pages 1--1, 2022.

\bibitem{DBLP:conf/ccs/ZhangDYZ14}
Mu~Zhang, Yue Duan, Heng Yin, and Zhiruo Zhao.
\newblock Semantics-aware android malware classification using weighted
  contextual {API} dependency graphs.
\newblock In {\em Proc. {SIGSAC}}, 2014.

\bibitem{DBLP:conf/ccs/ZhangZZDCZZY20}
Xiaohan Zhang, Yuan Zhang, Ming Zhong, Daizong Ding, Yinzhi Cao, Yukun Zhang,
  Mi~Zhang, and Min Yang.
\newblock Enhancing state-of-the-art classifiers with {API} semantics to detect
  evolved android malware.
\newblock In {\em Proc. {CCS}}, 2020.

\bibitem{DBLP:conf/iclr/ZhaoDS18}
Zhengli Zhao, Dheeru Dua, and Sameer Singh.
\newblock Generating natural adversarial examples.
\newblock In {\em Proc. {ICLR}}, 2018.

\bibitem{DBLP:journals/iotj/ZhouLLYSW22}
Xiaokang Zhou, Wei Liang, Weimin Li, Ke~Yan, Shohei Shimizu, and Kevin~I{-}Kai
  Wang.
\newblock Hierarchical adversarial attacks against graph-neural-network-based
  iot network intrusion detection system.
\newblock {\em {IEEE} Internet Things J.}, 9(12):9310--9319, 2022.

\bibitem{DBLP:conf/ndss/ZhouWZJ12}
Yajin Zhou, Zhi Wang, Wu~Zhou, and Xuxian Jiang.
\newblock Hey, you, get off of my market: Detecting malicious apps in official
  and alternative android markets.
\newblock In {\em Proc. {NDSS}}, 2012.

\bibitem{DBLP:conf/iclr/ZugnerG19}
Daniel Z{\"{u}}gner and Stephan G{\"{u}}nnemann.
\newblock Adversarial attacks on graph neural networks via meta learning.
\newblock In {\em Proc. {ICLR}}, 2019.

\end{thebibliography}

\section{Appendix}

\subsection{The limitations in callee selection} \label{sec:callees' limitation}
As shown in Section \ref{sec:adversarial manipulation method}, we choose the leaf nodes as candidate callees. However, not all leaf nodes can be chosen as callees. There exist two limitations:
\begin{itemize}
\item \textit{Access modifier}. Some leaf-node functions are not allowed to be invoked at all. Therefore, we only consider those leaf-node functions whose access modifier is \textit{public}. 
\item \textit{Parameter type}. The arguments of some leaf-node functions are the instances of classes. Under this situation, invoking these functions will incur instantiating a class, hence generating an unintended edge. To avoid this problem, we propose to choose the leaf-node functions whose arguments are void or belong to the category of primitive data types (e.g., int and short) and the \text{String} class. 
\end{itemize}

\subsection{Theoretical Analyses for our method} \label{appendix: theoretical analyses}

Our method {\ourtool} utilizes the algorithm {\ouralgorithm} to find the desired perturbation for a given malware sample. Since {\ouralgorithm} is an evolutionary algorithm, how to mitigate premature convergence is an important issue. Here, premature convergence or prematurity is a common phenomenon that leads an evolutionary algorithm to converge quickly to a local optimum. For evolutionary algorithms, prematurity is often caused by the lack of gene diversity.

 In the following, we analyze how multiple populations introduced in {\ouralgorithm} mitigate the problem of premature convergence.

Due to the introduction of multiple populations, there exist a local optimal solution in each {population}. We define this locally optimal solution as $x^{*}_{p}$, where $p=1,2,...,l$ is the index of the population.

Then, the individuals that can achieve  the local optimal solutions with the {\ouralgorithm} $G$ are termed as:
\begin{equation}
	A_{p}^{*}=\left\{x \in A: G(x)=x^{*}_{p}\right\}
\end{equation}
where $A$ is the solution space.

Then, the probability {that an} individual $x\in A$ {belongs} to set $A_{p}^{*}$ can be represented as $\theta_{p}=P\left(A_{p}^{*}\right)$. It is clear that $\theta_{p}>0$ for $p=1, \ldots, l$ and $\sum_{p=1}^{l} \theta_{p}=1$.

The size of the set  $A_{p}^{*}$ can be termed as $n_{p}$.
According to the definition, we have $n_{p} \geq 0\ (p=1, \ldots, l)$, the random vector $\left(N_{1}, \ldots, N_{l}\right)$ follows the multinomial distribution and $\sum_{p=1}^{l} N_{p}=N$.
\begin{equation}
	\operatorname{Pr}\left\{n_{1}=N_{1}, \ldots, n_{l}=N_{l}\right\}=\left(\begin{array}{c}
		N \\
		N_{1}, \ldots, N_{l}
	\end{array}\right) \theta_{1}^{N_{1}} \ldots \theta_{l}^{N_{l}}
\end{equation}
where
\begin{equation}
	\quad\left(\begin{array}{c}
		N \\
		N_{1}, \ldots, N_{l}
	\end{array}\right)=\frac{N !}{N_{1} ! \ldots N_{l} !}, \quad N_{p} \geq 0 \quad(p=1, \ldots, l)
\end{equation}

We define $W$ as the number of  locally  optimal solutions found by {\ouralgorithm}. Then the probability of   $l$ locally  optimal solutions being found can be termed  as 
\begin{equation}
	\operatorname{Pr}\{W=l \mid \theta\}=\sum_{N_{1}+\cdots+N_{l}=N} \left(\begin{array}{c}
		N \\
		N_{1}, \ldots, N_{l}
	\end{array}\right) \theta_{{1}}^{N_{1}} \ldots \theta_{{l}}^{N_{l}} 
\end{equation}
where
\begin{equation}
	\theta=\left(\theta_{1}, \ldots, \theta_{l}\right).
\end{equation}

For the sake of analyzing the limit, we define
\begin{equation}	
	\delta=\min \left\{\theta_{1}, \ldots, \theta_{l}\right\} \leq 1 / l
\end{equation}

Then we have
\begin{equation}
	\begin{aligned}
		\operatorname{Pr}\{W=l \mid \theta\} & \geq \sum_{N_{1}+\ldots+N_{l}=N}\left(\begin{array}{c}
			N \\
			N_{1}, \ldots, N_{l}
		\end{array}\right) \delta^{N} \\
		&=(\delta l)^{N} \operatorname{Pr}\left\{W=l \mid\left(\frac{1}{l}, \ldots, \frac{1}{l}\right)\right\}
	\end{aligned}
\end{equation}

For any $l$ and $\theta$, we can find the least evaluation number $n^*$ such that for any given $\gamma \in(0,1)$, we will have $\operatorname{Pr}\{W=l \mid \theta\} \geq \gamma$ 
for all $n \geq n^{*}$.
Finding $n^{*}=n^{*}(\gamma, \theta)$ is the problem of finding  the 
(minimal) number of points in $A$ such that the probability that
all local minimizers will be found is at least $\gamma$.

We analyze the extreme cases that $\theta^{*}=\left(l^{-1}, \ldots, l^{-1}\right)$. Hence the problem of finding $n^{*}(\gamma, \theta)$ is reduced to that of finding $n^{*}\left(\gamma, \theta^{*}\right)$. For a large $N$, $n^{*}(\gamma, \theta)$ can be approximated as 

\begin{equation}
	\begin{aligned}
		\operatorname{Pr}\left\{W=l \mid \theta^{*}\right\}&=l^{-N}  \sum_{N_{1}+\cdots+N_{l}=N}\left(\begin{array}{c}
			N \\
			N_{1}, \ldots, N_{l}
		\end{array}\right) \\
		&= \sum_{p=0}^{l}(-1)^{p}\left(\begin{array}{c}
			l \\
			p
		\end{array}\right)(1-p / l)^{N} \\
		& \sim \exp \{-l \exp \{-N / l\}\}, \quad N \rightarrow \infty
	\end{aligned}
\end{equation}

By solving the equation $\exp (-l \exp (-N / l))=\gamma$ with respect to $N$, we obtain
the approximation

\begin{equation}
	n^{*}\left(\gamma, \theta^{*}\right) \simeq l \ln l+l \ln (-\ln \gamma) \label{muti-p}
\end{equation}


With Eq. \eqref{muti-p}, we analyze the relationship between the number of required queries and the population number as follows. We can see that multiple populations ( i.e., $l>1$ ) help to slow down the convergence rate of the algorithm. As we all know, prematurity is a common  phenomenon in which an evolutionary algorithm early converges to a poor local optimum. However, {\ouralgorithm} begins its search in multi start $l$ which  makes the algorithm can find a better solution with a higher probability. Our algorithm effectively relieves this problem by introducing multiple populations, and prevents the algorithm from wasting many efforts on repeatedly finding the same local optimum.

\subsection{ {Implementation details and an instance of the smali code}}
\label{appendix:smali}

	\begin{figure}[!h]
		\centering
		\includegraphics[scale=0.6]{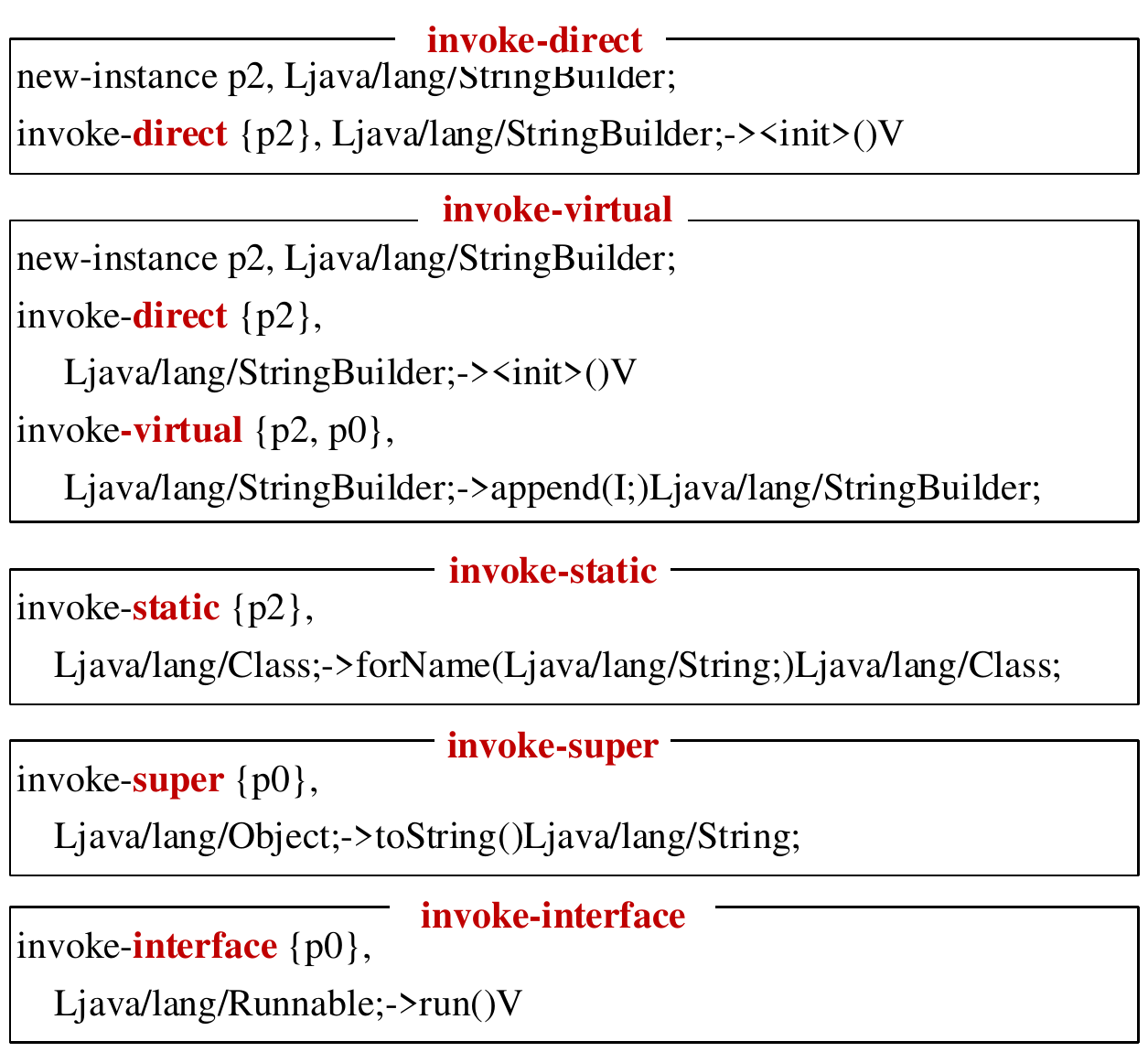}
		\caption{The examples of smali code with different invocation types.}
		\label{fig:direct}
	\end{figure}

 {
In this section, we first provide the implementation details of the transformation from the generator’s output to the perturbation on the malware samples. Then we give an instance of the samli code.}
 {The output of the generator is pairs of caller-callee functions. There are three steps to implement output-to-perturbation transformation. First, for every function pair, we find the smali file related to the selected caller, according to the latter's full name. }
 {Second, we insert statements into the smali file to implement a try-catch trap. Here we can use five types of function invocation, including invoke-direct, invoke-virtual, invoke-static, invoke-super and invoke-interface. Different invocation types require different smali manipulation. Fig. \ref{fig:direct} shows an example for every invocation type.  }
 {Third, we use \textit{Apktool} to rebuild the modified smali files to APK file. The above operations are automatically conducted by a Python script. }

\begin{figure}[!h]
	\centering
	\includegraphics[scale=0.45]{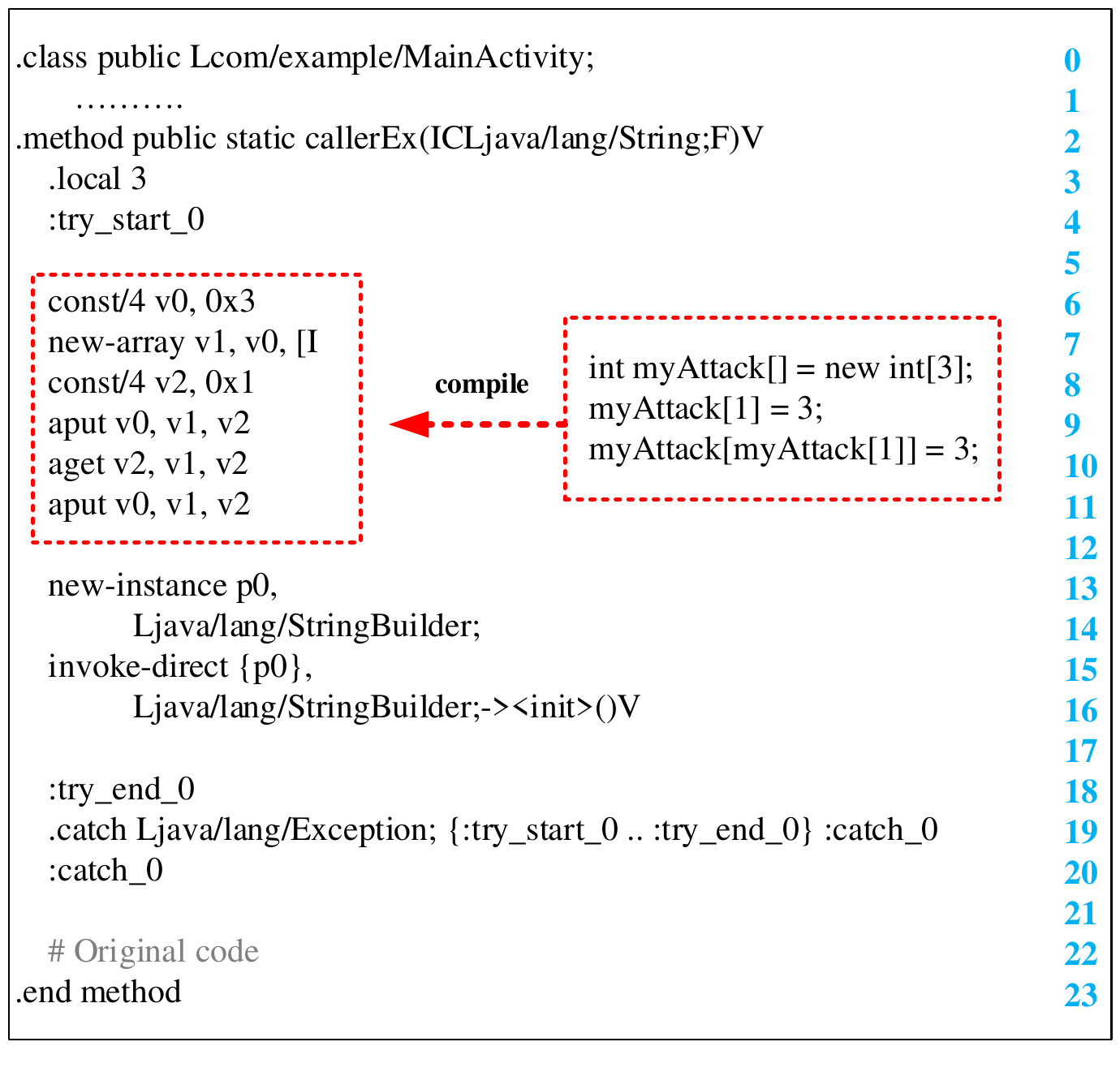}
	\caption{An instance of the smali code.}
	\label{fig:smali}
\end{figure}

\begin{figure}[!h]
	\centering
	\includegraphics[scale=1]{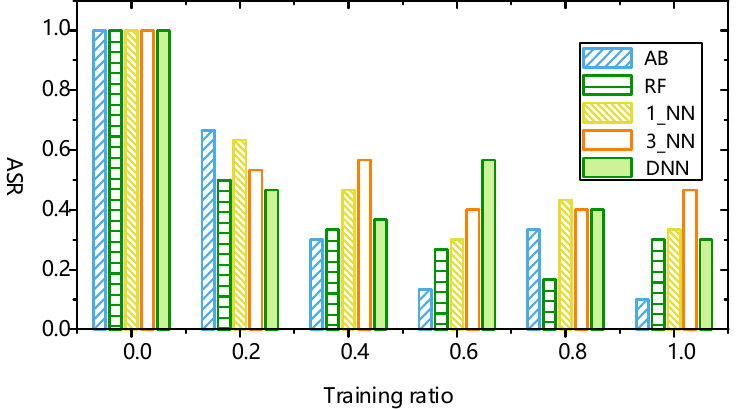}
	\caption{{Attack success rate after retraining.}}
	\label{fig:retraining}
\end{figure}

 \begin{figure}[!h]
	\centering
	\includegraphics[scale=0.65]{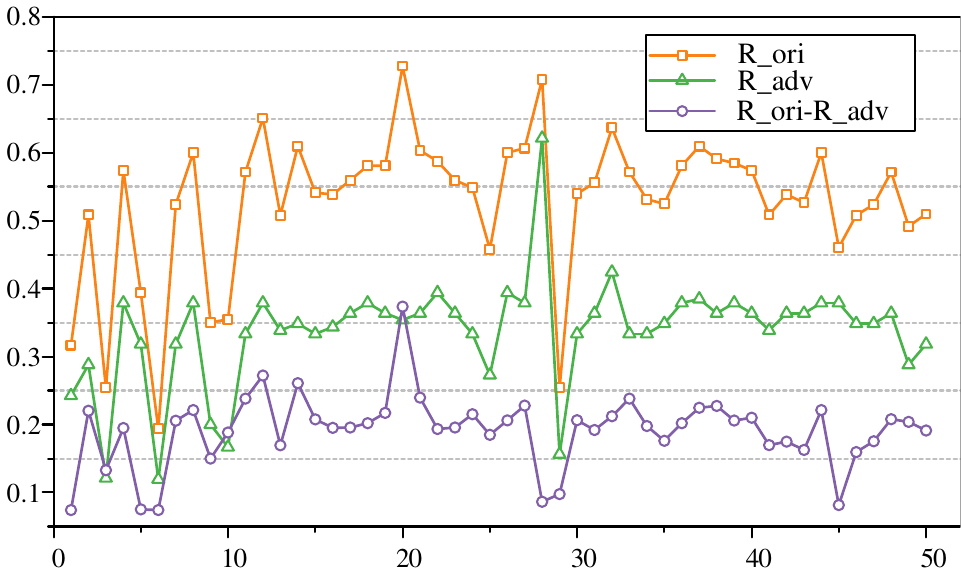}
	\caption{ {The detection success ratio on VirusTotal}.}
	\label{fig:ratio}
\end{figure}

 \begin{figure*}[!h]
	\centering
	\includegraphics[scale=0.9]{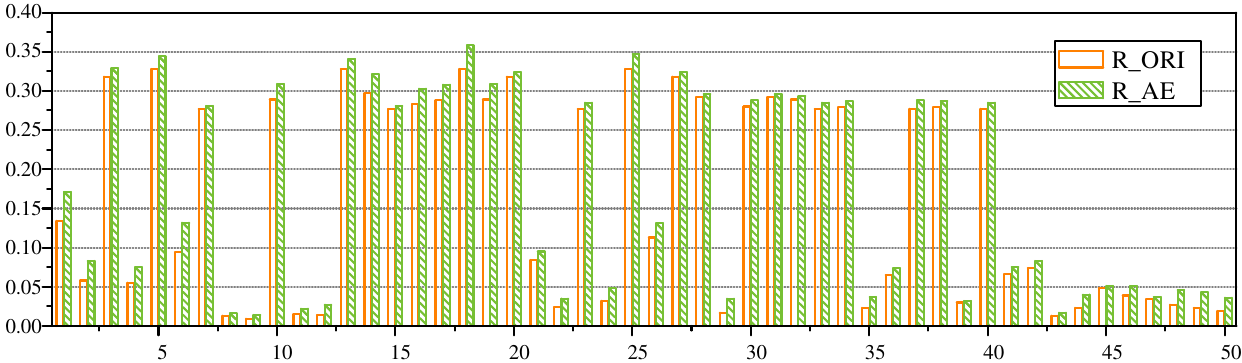}
	\caption{ {The number of try-catch blocks before and after {\ourtool} attack.}}
	\label{fig:try_catch}
\end{figure*}


To show how {\ourtool}  manipulates the smali code, we supply a  practical manipulation instance in Fig. \ref{fig:smali}. From line 6 to line 11, {we can find} a runtime exception. To be specific, we initialize an array with length 3 and employ an opaque method to visit the 4-th element of this array. Then it will throw an exception \textit{java.lang.ArrayIndexOutOfBoundsException} and skip the inserted callee functions. In this way, our method can effectively insert calls and preserve the malware's original functionality.


{It is worth noting that the statements added into a try block are not fixed. Hence {\ourtool} can resist the whitelist-based defense. For example, suppose we want to trigger the exception of \textit{IndexOutOfBoundsException} by inducing array access violation. For this purpose, we access the array index that exceeds the array length. {\ourtool} can generate countless variable names and variable values for such an array index. Therefore, it is impossible to build a white list to rule out the statements added by {\ourtool}.}

\subsection{\textbf{Dataset in our experiments}} \label{appendix: dataset}
{ Our dataset includes 44375 Android APKs released from 2010 to 2020, which are collected from AndroZoo, FalDroid,and Drebin. Table \ref{tab:dataset} gives the source, count,and years of APKs in our dataset. }

\begin{table}[H]
	\caption{{Dataset used in our experiments.}}
	\centering
	\label{tab:dataset}
\scalebox{1}{%
\begin{tabular}{c|ccc}
\hline
\textbf{Source} & Label & Years & Count \\ \hline
\multirow{2}{*}{Androdzoo} & \cellcolor{cyan!60!gray!10}Benign & \cellcolor{cyan!60!gray!10}2010-2020 & \cellcolor{cyan!60!gray!10}21399 \\
 & Malicious & 2015-2020 & 9668 \\ \hline
FalDroid & \cellcolor{cyan!60!gray!10}Malicious & \cellcolor{cyan!60!gray!10}2013-2014 & \cellcolor{cyan!60!gray!10}8407 \\ \hline
Drebin & Malicious & 2010-2012 & 4900 \\ \hline
Total & \cellcolor{cyan!60!gray!10}- & \cellcolor{cyan!60!gray!10}2010-2020 & \cellcolor{cyan!60!gray!10}44374 \\ \hline
\end{tabular}%
}
\end{table}

\subsection{Resistance to adversarial retraining.}
\label{app:ar}

Adversarial retraining is regarded as the most effective defense method against AE attacks. In this section, we test {\ourtool} with adversarial retraining. We randomly select 100 adversarial examples that are generated by {\ourtool} and can deceive target systems. We divide these adversarial examples into a training and a test set. {Under various training sample proportions, we retrain the target classifier in order to evaluate ASR on the test set.}


{Our results are given in Fig. \ref{fig:retraining}, whose vertical axis is the ASR of {\ourtool} and the horizontal axis is the proportion of AEs used in adversarial retraining. Not surprisingly, the ASR decreases with the increase of the AEs adopted by adversarial retraining. When the ratio exceeds 40\%, adversarial retraining becomes effective in resisting {\ourtool}. 
In practice, however, it is extremely difficult to collect sufficient adversarial examples for adversarial retraining. On the other hand, it is also noted that aided by {\ourtool}, model owners can improve their models’ defense capability with adversarial retraining.
}

\subsection{  {Attack performance on VirusTotal}}
\label{app:virustotal}

We  evaluate the  performance of {\ourtool} on  VirusTotal. 
 To be specific, we use {\ourtool} to generate AEs (adversarial examples) through querying the MaMadroid detector, and upload them to VirusTotal for malware detection.  
 VirusTotal uses about 60 malware detection methods unknown to us. We then record the ratio of the successful detection methods to all the methods, denoted by R\_adv. For comparison, we also conduct the same setting for the original sample, and the corresponding ratio is termed as R\_ori. 
 What's more, we calculate the difference between the R\_adv and R\_ori, which termed as R\_ori - R\_adv.
 The results are shown in Fig. \ref{fig:ratio}.  The horizontal axis of this figure shows different APKs, and the vertical axis gives the ratios R\_adv (denoted by the red line) and R\_ori (denoted by the blue line). The yellow line shows the decreasing ratio of the successful detection methods. It can be seen that {\ourtool} can effectively reduce the probability of malware being detected, owing to the transferability of AEs \cite{DBLP:conf/iclr/LiuCLS17}. It is worth noting that this attack effect is achieved under the scenario where no queries are conducted and no prior knowledge about detection methods can be obtained.

\subsection{  {The number of added try-catch blocks}}\label{app:trycatch}

 {Since {\ourtool} inserts try-catch blocks into malware code, a defender may choose to detect it through judging whether the number of try-catch blocks exceeds a predetermined threshold. However, it is difficult to find an appropriate threshold for all APKs. Without such a threshold, this defense method may cause a high false positive or false negative rate.}

 {To verify it, we record the ratio of try-catch block number to the function-calls number in 50 malicious APKs and the corresponding adversarially perturbed APKs, termed as R\_ORI and R\_AE, respectively. The results are shown in Fig. \ref{fig:try_catch}.
The horizontal axis of this figure shows the IDs of these APKs, and the vertical axis gives the count ratio of try-catch blocks. The orange and green bars mean the original APK and the  corresponding modified APK, respectively. We can draw two conclusions from this figure. First, the number of try-catch blocks added by our method is relatively small compared to that of existing try-catch blocks. Therefore it is hard to find a threshold to 
 clearly distinguish the original APK and the perturbed APK. Second, the number of try-catch blocks drastically fluctuates among various APKs. Thus, it is also difficult to set a fixed threshold  for all APKs. }

\end{document}